\RequirePackage{lineno} 
\documentclass[aps,prd,twocolumn,superscriptaddress,preprintnumbers,amsmath,amssymb,tightenlines,floatfix]{revtex4-1}
\usepackage{graphicx} 
\usepackage{dcolumn}  
\usepackage{rotating}
\usepackage{epstopdf}
\usepackage{hyperref}
\graphicspath{{ps}}

\begin{document}


\preprint{\vbox{ 
    \hbox{Belle DRAFT {\it 2017-13}}
    \hbox{KEK DRAFT {\it 2017-09}}
}}

\title{ \quad\\[1.0cm]Invariant-mass and fractional-energy dependence of inclusive production of di-hadrons in $e^+e^-$ annihilation at $\sqrt{s}=$ 10.58 GeV }
\noaffiliation
\affiliation{University of the Basque Country UPV/EHU, 48080 Bilbao}
\affiliation{Beihang University, Beijing 100191}
\affiliation{Budker Institute of Nuclear Physics SB RAS, Novosibirsk 630090}
\affiliation{Faculty of Mathematics and Physics, Charles University, 121 16 Prague}
\affiliation{Chonnam National University, Kwangju 660-701}
\affiliation{University of Cincinnati, Cincinnati, Ohio 45221}
\affiliation{Deutsches Elektronen--Synchrotron, 22607 Hamburg}
\affiliation{University of Florida, Gainesville, Florida 32611}
\affiliation{Justus-Liebig-Universit\"at Gie\ss{}en, 35392 Gie\ss{}en}
\affiliation{SOKENDAI (The Graduate University for Advanced Studies), Hayama 240-0193}
\affiliation{Gyeongsang National University, Chinju 660-701}
\affiliation{Hanyang University, Seoul 133-791}
\affiliation{University of Hawaii, Honolulu, Hawaii 96822}
\affiliation{High Energy Accelerator Research Organization (KEK), Tsukuba 305-0801}
\affiliation{J-PARC Branch, KEK Theory Center, High Energy Accelerator Research Organization (KEK), Tsukuba 305-0801}
\affiliation{IKERBASQUE, Basque Foundation for Science, 48013 Bilbao}
\affiliation{Indian Institute of Science Education and Research Mohali, SAS Nagar, 140306}
\affiliation{Indian Institute of Technology Bhubaneswar, Satya Nagar 751007}
\affiliation{Indian Institute of Technology Guwahati, Assam 781039}
\affiliation{Indian Institute of Technology Madras, Chennai 600036}
\affiliation{Indiana University, Bloomington, Indiana 47408}
\affiliation{Institute of High Energy Physics, Chinese Academy of Sciences, Beijing 100049}
\affiliation{Institute of High Energy Physics, Vienna 1050}
\affiliation{Institute for High Energy Physics, Protvino 142281}
\affiliation{INFN - Sezione di Napoli, 80126 Napoli}
\affiliation{INFN - Sezione di Torino, 10125 Torino}
\affiliation{Advanced Science Research Center, Japan Atomic Energy Agency, Naka 319-1195}
\affiliation{J. Stefan Institute, 1000 Ljubljana}
\affiliation{Kanagawa University, Yokohama 221-8686}
\affiliation{Institut f\"ur Experimentelle Kernphysik, Karlsruher Institut f\"ur Technologie, 76131 Karlsruhe}
\affiliation{Kennesaw State University, Kennesaw, Georgia 30144}
\affiliation{King Abdulaziz City for Science and Technology, Riyadh 11442}
\affiliation{Department of Physics, Faculty of Science, King Abdulaziz University, Jeddah 21589}
\affiliation{Korea Institute of Science and Technology Information, Daejeon 305-806}
\affiliation{Korea University, Seoul 136-713}
\affiliation{Kyoto University, Kyoto 606-8502}
\affiliation{Kyungpook National University, Daegu 702-701}
\affiliation{\'Ecole Polytechnique F\'ed\'erale de Lausanne (EPFL), Lausanne 1015}
\affiliation{P.N. Lebedev Physical Institute of the Russian Academy of Sciences, Moscow 119991}
\affiliation{Faculty of Mathematics and Physics, University of Ljubljana, 1000 Ljubljana}
\affiliation{Ludwig Maximilians University, 80539 Munich}
\affiliation{Luther College, Decorah, Iowa 52101}
\affiliation{University of Maribor, 2000 Maribor}
\affiliation{Max-Planck-Institut f\"ur Physik, 80805 M\"unchen}
\affiliation{School of Physics, University of Melbourne, Victoria 3010}
\affiliation{University of Miyazaki, Miyazaki 889-2192}
\affiliation{Moscow Physical Engineering Institute, Moscow 115409}
\affiliation{Moscow Institute of Physics and Technology, Moscow Region 141700}
\affiliation{Graduate School of Science, Nagoya University, Nagoya 464-8602}
\affiliation{Kobayashi-Maskawa Institute, Nagoya University, Nagoya 464-8602}
\affiliation{Nara Women's University, Nara 630-8506}
\affiliation{National Central University, Chung-li 32054}
\affiliation{National United University, Miao Li 36003}
\affiliation{Department of Physics, National Taiwan University, Taipei 10617}
\affiliation{H. Niewodniczanski Institute of Nuclear Physics, Krakow 31-342}
\affiliation{Nippon Dental University, Niigata 951-8580}
\affiliation{Niigata University, Niigata 950-2181}
\affiliation{Novosibirsk State University, Novosibirsk 630090}
\affiliation{Osaka City University, Osaka 558-8585}
\affiliation{Pacific Northwest National Laboratory, Richland, Washington 99352}
\affiliation{University of Pittsburgh, Pittsburgh, Pennsylvania 15260}
\affiliation{Theoretical Research Division, Nishina Center, RIKEN, Saitama 351-0198}
\affiliation{RIKEN BNL Research Center, Upton, New York 11973}
\affiliation{University of Science and Technology of China, Hefei 230026}
\affiliation{Showa Pharmaceutical University, Tokyo 194-8543}
\affiliation{Soongsil University, Seoul 156-743}
\affiliation{Stefan Meyer Institute for Subatomic Physics, Vienna 1090}
\affiliation{Sungkyunkwan University, Suwon 440-746}
\affiliation{School of Physics, University of Sydney, New South Wales 2006}
\affiliation{Department of Physics, Faculty of Science, University of Tabuk, Tabuk 71451}
\affiliation{Excellence Cluster Universe, Technische Universit\"at M\"unchen, 85748 Garching}
\affiliation{Department of Physics, Technische Universit\"at M\"unchen, 85748 Garching}
\affiliation{Toho University, Funabashi 274-8510}
\affiliation{Department of Physics, Tohoku University, Sendai 980-8578}
\affiliation{Earthquake Research Institute, University of Tokyo, Tokyo 113-0032}
\affiliation{Department of Physics, University of Tokyo, Tokyo 113-0033}
\affiliation{Tokyo Institute of Technology, Tokyo 152-8550}
\affiliation{Tokyo Metropolitan University, Tokyo 192-0397}
\affiliation{University of Torino, 10124 Torino}
\affiliation{Virginia Polytechnic Institute and State University, Blacksburg, Virginia 24061}
\affiliation{Wayne State University, Detroit, Michigan 48202}
\affiliation{Yamagata University, Yamagata 990-8560}
\affiliation{Yonsei University, Seoul 120-749}
  \author{R.~Seidl}\affiliation{RIKEN BNL Research Center, Upton, New York 11973} 
  \author{I.~Adachi}\affiliation{High Energy Accelerator Research Organization (KEK), Tsukuba 305-0801}\affiliation{SOKENDAI (The Graduate University for Advanced Studies), Hayama 240-0193} 
  \author{H.~Aihara}\affiliation{Department of Physics, University of Tokyo, Tokyo 113-0033} 
  \author{S.~Al~Said}\affiliation{Department of Physics, Faculty of Science, University of Tabuk, Tabuk 71451}\affiliation{Department of Physics, Faculty of Science, King Abdulaziz University, Jeddah 21589} 
  \author{D.~M.~Asner}\affiliation{Pacific Northwest National Laboratory, Richland, Washington 99352} 
  \author{T.~Aushev}\affiliation{Moscow Institute of Physics and Technology, Moscow Region 141700} 
 \author{R.~Ayad}\affiliation{Department of Physics, Faculty of Science, University of Tabuk, Tabuk 71451} 
  \author{I.~Badhrees}\affiliation{Department of Physics, Faculty of Science, University of Tabuk, Tabuk 71451}\affiliation{King Abdulaziz City for Science and Technology, Riyadh 11442} 
  \author{A.~M.~Bakich}\affiliation{School of Physics, University of Sydney, New South Wales 2006} 
  \author{V.~Bansal}\affiliation{Pacific Northwest National Laboratory, Richland, Washington 99352} 
% ...
\collaboration{The Belle Collaboration}

\noaffiliation
\begin{abstract}
The inclusive cross sections for di-hadrons of charged pions and kaons ($e^+e^- \rightarrow hhX$) in electron-positron annihilation are reported. They are obtained as a function of the total fractional energy and invariant mass for any di-hadron combination in the same hemisphere as defined by the thrust event-shape variable and its axis. Since same-hemisphere di-hadrons can be assumed to originate predominantly from the same initial parton, di-hadron fragmentation functions are probed. These di-hadron fragmentation functions are needed as an unpolarized baseline in order to quantitatively understand related spin-dependent measurements in other processes and to apply them to the extraction of quark transversity distribution functions in the nucleon.  
The di-hadron cross sections are obtained from a $655\,{\rm fb}^{-1}$ data sample collected at or near the $\Upsilon(4S)$ resonance with the Belle detector at the KEKB asymmetric-energy $e^+ e^-$
collider.
\end{abstract}


\maketitle

{\renewcommand{\thefootnote}{\fnsymbol{footnote}}}
\setcounter{footnote}{0}

Fragmentation functions allow us to understand the transition of asymptotically free partons into several confined hadrons. They cannot be calculated from first principles and thus need to be extracted experimentally. One of the main ways of obtaining them is via cross section or multiplicity measurements in electron-positron annihilation where no hadrons are present in the initial state. For many processes, factorization is assumed or proven to certain orders of the strong coupling and fragmentation functions as well as parton distribution functions are considered universal. Because of this universality, such functions extracted in one process can be applied to another process. As such, the knowledge of fragmentation functions is, for example, used to extract various spin-dependent parton distribution functions in polarized semi-inclusive deep-inelastic scattering (SIDIS) and polarized hadron collisions. 
In particular, the extraction of the chiral-odd transversity distribution functions \cite{ralstonsoper} and their related tensor charges relies so far entirely on transverse spin dependent fragmentation functions. 

The Belle experiment was the first to provide asymmetries \cite{bellecollins} related to the single-hadron Collins fragmentation function \cite{Collins:1992kk}. These asymmetries rely on an explicit transverse-momentum dependence of fragmentation functions. The Collins fragmentation function describes a correlation between the direction of an outgoing transversely polarized quark, its spin orientation and the azimuthal distribution of final-state hadrons, and serves as a transverse-spin analyzer. Collins asymmetries were extracted for pions and kaons in several SIDIS measurements so far \cite{hermes,hermes10,compasscollins,compass10,halla}, where they are convolved with the transversity distributions of interest, and more recently in proton-proton collisions for pions \cite{starcollins}. The corresponding Collins fragmentation measurements were obtained in various electron-positron annihilation experiments for pions \cite{bellecollins,babarcollins,bes3} and recently also kaons \cite{babark} based on the description of Ref.~\cite{boercollins}. Some of these measurements have already been included in global transversity extractions \cite{torino,kang,alexeik,torino15}.

An alternative way of accessing quark transversity is via di-hadron fragmentation functions \cite{Collins:1993kq,jaffe,Radici:2001na}. This has the advantage of being based on collinear factorization. Here, too, Belle has provided the corresponding asymmetries related to the polarized fragmentation functions \cite{belleiff} following the description in Ref.~\cite{boer}. They were used with the SIDIS measurements \cite{hermesdiff,compassdiff} in a global analysis \cite{radici} to extract transversity in a collinear approach. The relevant measurements from proton-proton collisions \cite{stariff} were not part of these global fits but appear to be consistent with them \cite{Radici:2016lam}.

In both approaches of transversity extraction, several assumptions are made due to the lack of sufficient measurements. In the Collins-based extractions, the explicit transverse-momentum dependence was until recently unknown and is still poorly constrained. In the di-hadron based extractions, the corresponding unpolarized di-hadron fragmentation functions were not available so far and theorists used Monte Carlo (MC) simulations to estimate those. This publication provides the unpolarized baseline for the measurements related to the spin-dependent di-hadron fragmentation functions. 

In a previous publication \cite{dihadronprd1}, the focus was on two-hadron cross sections differential in their individual fractional energies $z_1=2E_{h_1}/\sqrt{s}$ and (likewise) $z_2$. In this description, the two-hadron production can be described by di-hadron fragmentation functions (DiFF), generally introduced in Ref.~\cite{Bianconi:1999cd} and specifically for these DiFFs in Ref.~\cite{deFlorian:2003cg} and based on the formalism developed in Ref.~\cite{diff}. DGLAP \cite{dglap} evolution for DiFFs was also introduced previously \cite{diffevo1,diffevo2}. Recently, this theoretical work has been applied also to DiFFs depending explicitly on the combined fractional energy $z=2E_{h_1h_2}/\sqrt{s}$ and invariant mass $m_{h_1h_2}$ of the hadrons, instead of the hadrons' individual fractional energies, and including evolution as summarized in Ref.~\cite{Ceccopieri:2007ip}. It is in this description that the SIDIS measurements and the Belle asymmetries were performed and, here, we report the corresponding cross sections differential in these two variables to provide the unpolarized baseline.
 
\begin{figure}[ht]
\begin{center}
\includegraphics[width=8cm]{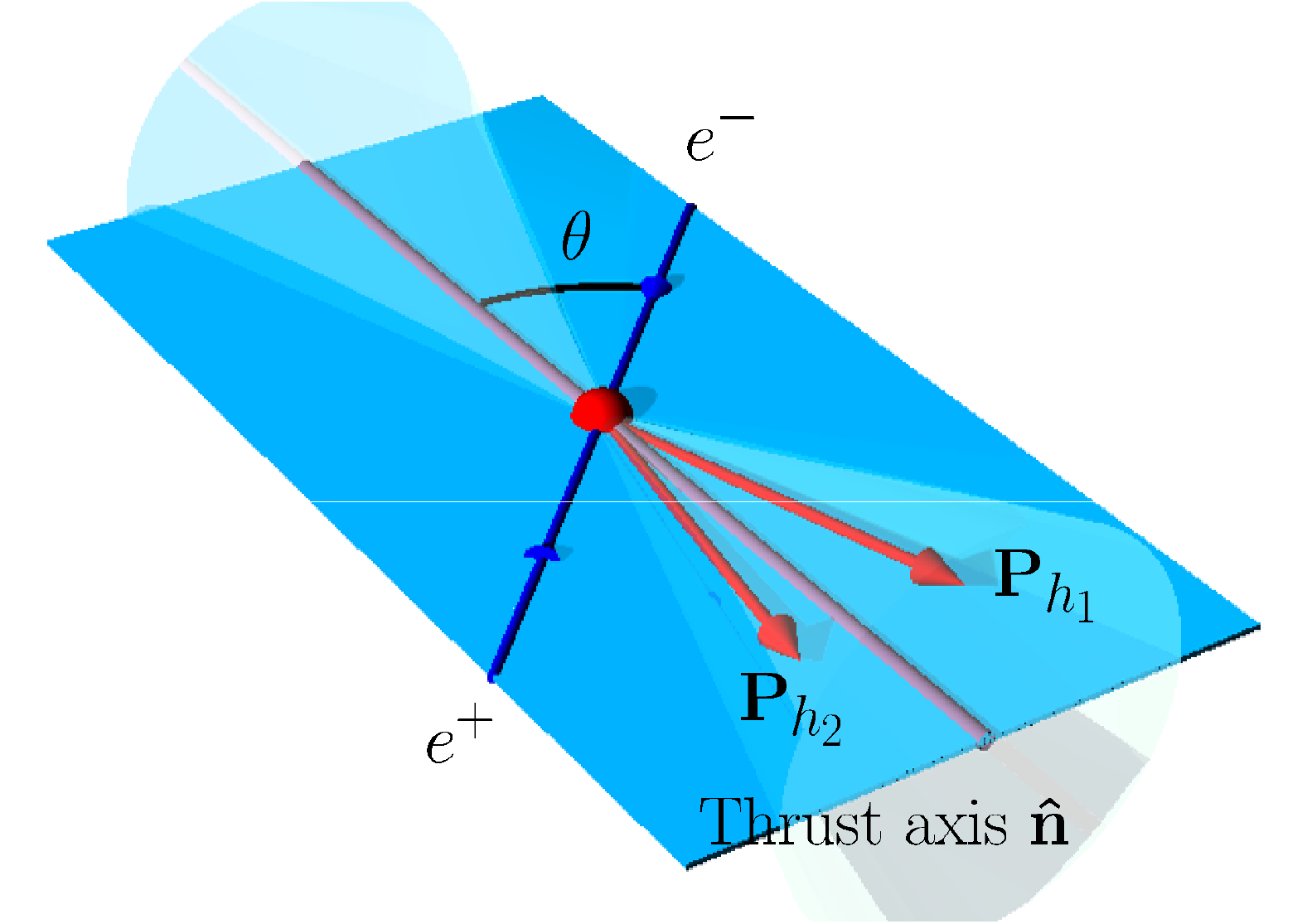} 
\caption{\label{fig:(hh)x} Illustration of di-hadron production where the final-state hadrons with momenta $P_{h_1}$ and $P_{h_2}$ in the CMS are depicted as red arrows, the incoming leptons as blue arrows, and the event plane--spanned by leptons and thrust axis--is depicted as the light-blue plane. In this case, both hadrons are found in the same hemisphere as defined by the thrust axis, and generally out of the plane, as indicated by the cones.}
\end{center}
\end{figure}

The cross section at leading order in the strong coupling can be described as 
\begin{align}
&\frac{{\mathrm d}^2\sigma(e^+e^-\rightarrow h_1h_2X)}{{\mathrm d}z{\mathrm d}m_{h_1h_2}} \propto&\nonumber \\&\sum_q e_q^2 \left( D_{1,q}^{h_1h_2}(z,m_{h_1h_2}) +  D_{1,\overline{q}}^{h_1h_2}(z,m_{h_1h_2})\right),&
\label{eq:eehhx}
\end{align}
where it is assumed that both hadrons emerge from the same (anti)quark, $q$, and the scale dependence has been dropped for brevity. The assumption that hadrons detected in the same hemisphere, as illustrated in Fig.~\ref{fig:(hh)x}, originate from the same initial parton is supported by the results of Ref.~\cite{dihadronprd1}. To define the hemispheres, a selection of thrust axis and thrust value is required. The thrust axis $\hat{\mathbf{n}}$ maximizes the thrust $T$ \cite{thrust}:
\begin{equation}
T \stackrel{\mathrm{max}}{=} \frac{ \sum_h|\mathbf{P}^{\mathrm{CMS}}_h\cdot\mathbf{\hat{n}}|}{ \sum_h|\mathbf{P}^{\mathrm{CMS}}_h|}\quad.
\end{equation}
The sum extends over all detected particles, and $\mathbf{P}^{\mathrm{CMS}}_h$ denotes the three-momentum of particle $h$ in the ($e^+e^-$) center-of-mass system (CMS).

The cross sections for the inclusive production of di-hadrons of charged pions and kaons in the same hemisphere as a function of their fractional energy $z$ and invariant mass $m_{h_1h_2}$ are presented in this paper. The cross sections are compared to various MC simulation tunes optimized for different collision systems and energies. Various resonances in the mass spectra and distinct features from multi-body or subsequent decays of resonances are identified with the help of MC simulations. Additionally, the di-hadron cross sections after a MC-based removal of all weak decays are presented.\par
The paper is organized as follows. The detector setup and reconstruction criteria are summarized in Section \ref{sec1}. In Section \ref{sec2}, the various corrections to get from the raw spectra to the final cross sections are discussed. In Section \ref{sec3}, the results are shown and compared to MC tunes. We conclude with a summary in Section \ref{sec4}. A short Appendix discusses the partial waves and the impact the selection criteria have on various partial-wave moments as well as the weak-decay-removed cross sections.

\section{Belle detector and data selection\label{sec1}}

This di-hadron cross section measurement is based on a data sample of $655\,{\rm fb}^{-1}$, 
collected with the Belle detector at the KEKB asymmetric-energy
$e^+e^-$ (3.5~GeV on 8~GeV) collider~\cite{KEKB}
operating at the $\Upsilon(4S)$ resonance (denoted as on-resonance) as well as 60 MeV below for comparison (denoted as continuum).

The Belle detector is a large-solid-angle magnetic
spectrometer that consists of a silicon vertex detector (SVD),
a 50-layer central drift chamber, an array of
aerogel threshold Cherenkov counters,  
a barrel-like arrangement of time-of-flight
scintillation counters, and an electromagnetic calorimeter
comprising CsI(Tl) crystals located inside 
a superconducting solenoid coil that provides a 1.5~T
magnetic field.  An iron flux-return located outside of
the magnet coil is instrumented to detect $K_L^0$ mesons and to identify
muons.  The detector
is described in detail elsewhere~\cite{Belle}.
Two inner detector configurations were used. A 2.0 cm beampipe with 1 mm thickness 
and a 3-layer SVD were used for the first sample
of $97\,{\rm fb}^{-1}$, while a 1.5 cm beampipe, a 4-layer
SVD and a small-cell inner drift chamber were used to record  
the remaining $558\,{\rm fb}^{-1}$ \cite{svd2}.  

The primary light ($uds$)- and charm-quark simulations used in this analysis were generated using {\sc Pythia}6.2 \cite{pythia}, embedded into the {\sc EvtGen} \cite{evtgen} framework, followed by a {\sc Geant}3 \cite{geant} simulation of the detector response. The various MC samples were produced separately for light and charm quarks. For comparisons of data with generator-level MC simulations, weak decays, which normally are handled in {\sc Geant}, were allowed to decay in {\sc EvtGen}. In addition, we generated charged and neutral $B$ meson pairs from $\Upsilon(4S)$ decays in {\sc EvtGen}, $\tau$ pair events with the {\sc KKMC} \cite{taumc} generator and the {\sc Tauola} \cite{tauola} decay package, and other events with either {\sc Pythia} or dedicated generators \cite{aafh}. 
\subsection{Event and track selection}
Events with at least three reconstructed charged tracks must have a visible energy $E_{\mathrm{vis}}$ of all charged tracks and neutral clusters above 7 GeV (to remove $\tau$ pair events) and either a heavy jet mass (the greater of the two invariant masses of all particles in a hemisphere) above 1.8 GeV or a ratio of the heavy jet mass to visible energy above 0.25. 

Tracks must be within  $|dz|<$ 4 cm ($dr<$ 2 cm) of the interaction point along (perpendicular to) the positron beam axis. Each track must have at least three SVD hits and fall within the barrel and full particle-identification (PID) polar-angle acceptance of $-0.511  < \cos\theta_{\mathrm{lab}} < 0.842 $.
The fractional energy of each track must exceed 0.1. (Note that, in this paper, we study fragmentation functions for $z$ above 0.2.) This initial fractional-energy selection always takes the nominal hadron mass as given by the PID information into account. 

All two-hadron combinations were selected if found in the same hemisphere, as depicted in Fig.~\ref{fig:(hh)x}, where the hemispheres are defined by the plane perpendicular to the thrust axis. The thrust itself must satisfy $T>0.8$. The thrust axis needs to be within the barrel acceptance $|\hat{n}_z| < 0.75$. All selection criteria are summarized in Table \ref{cuts}.

\begin{table}
\begin{center}
\caption{\label{cuts}All selection criteria for this analysis are summarized in this table. Most of the track selection criteria are also applied for all the particles (including trackless clusters) considered for the thrust calculation.}
\begin{tabular}{c c} 
\multicolumn{2}{c}{Experiment selection} \\ \hline
On-resonance &  655.28 fb$^{-1}$ \\
Continuum comparison & 89.56 fb$^{-1}$\\ \hline
\multicolumn{2}{c}{Event selection} \\ \hline
Visible energy & $E_{\mathrm{vis}} >7$ GeV \\
Thrust value & $T > 0.8$ \\
Central thrust & $|\hat{n}_z| < 0.75$ \\ \hline 
\multicolumn{2}{c}{Track selection}\\ \hline
SVD hits & $N_{SVD}\geq 3$ \\
Barrel acceptance & $-0.511< \cos\theta_{\mathrm{lab}}<0.842$ \\
Lab momentum & $0.5$ GeV $ < P_{\mathrm{lab}} < 8.0 $ GeV \\
Lab transv.~momentum/energy & $P_{\mathrm{lab},T} > 0.1 $ GeV \\
Vertex $z$ & $|dz|<$4 cm \\
Vertex radius & $dr<$2 cm \\
raw $z$ cut & $z > 0.1$ \\\hline
\multicolumn{2}{c}{PID selection}\\ \hline
Electron likelihood ratio & $e >0.85$ and $\mu<0.9$\\
$\mu$ likelihood ratio& $\mu>0.9$ \\
Pion & $K\pi < 0.6$ and $ \pi p > 0.2$ \\
Kaon & $K\pi > 0.6$ and $ K p > 0.2$ \\
Proton & $\pi p <0.2$ and $K p <0.2$ \\ \hline
\end{tabular}
\end{center}
\end{table}

\begin{figure*}[th]
\begin{center}
\includegraphics[width=0.95\textwidth]{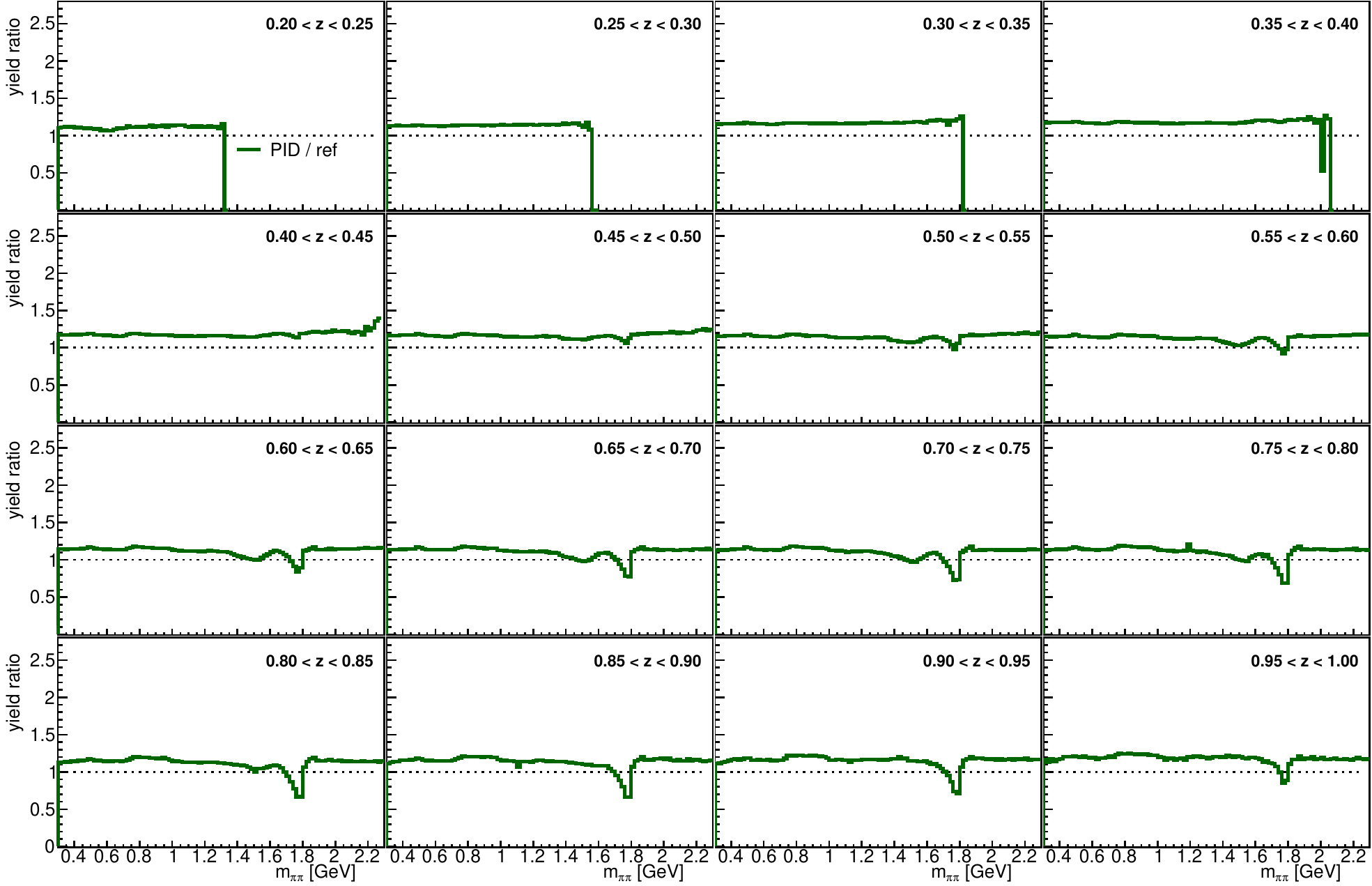} 
\caption{\label{fig:beforeafterpid} Ratio of yields after to before applying the PID correction for $\pi^+\pi^-$ pairs as a function of the invariant mass $m_{\pi\pi}$ in bins of $z$. Empty bins are visible where the yields become zero, especially for high-mass bins; kinematic limits are visible in the low-$z$ bins.}
\end{center}
\end{figure*}

\subsection{PID selection}
To apply the PID correction according to the PID efficiency matrices described in Ref.~\cite{martin}, the same selection criteria must be applied to define a charged track as a pion, kaon, proton, electron, or muon. The relevant information is determined from normalized likelihood ratios that are constructed from various detector responses.  
If the muon-hadron likelihood ratio is above 0.9, the track is identified as a muon. Otherwise, if the electron-hadron likelihood ratio is above 0.85, the track is identified as an electron. 
If neither of these applies, the track is identified as a kaon by a kaon-pion likelihood ratio above 0.6 and a kaon-proton likelihood ratio above 0.2. Pions are identified with the kaon-pion likelihood ratio below 0.6 and a pion-proton ratio above 0.2. Finally, protons are identified with the inverse proton ratios above with kaon-proton and pion-proton ratios below 0.2. While neither muons nor electrons are considered explicitly for the di-hadron analysis, they are retained as necessary contributors for the PID correction, wherein a certain fraction enter the pion, kaon and proton samples under study. These criteria are also summarized in Table \ref{cuts}.

The overall pion identification efficiencies are above 90\% at low laboratory momenta but drop to around 85\% at intermediate momenta where the majority are misidentified as kaons. Kaons have identification efficiencies above 90\% at low lab momenta and drop continuously to below 80\% at high lab momenta, with the majority of the misidentified kaons being reconstructed as pions or protons. Protons have similar reconstruction efficiencies, but drop to about 50\% above 3 GeV where they are almost as likely to be falsely identified as kaons. All particle identification efficiencies are flat as a function of the lab polar angle.  

\section{Di-hadron analysis and corrections\label{sec2}}
In the following subsections, the di-hadron yields are extracted and, successively, the various corrections and the corresponding systematic uncertainties are applied to arrive at the di-hadron differential cross sections ${\mathrm d}^2\sigma(e^+e^-\rightarrow h_1h_2X)/{\mathrm d}z{\mathrm d}m_{h_1h_2}$. 
\subsection{Binning and cross section extraction}
For the di-hadron cross sections, a ($z$, $m_{h_1h_2}$) binning is used as it is most relevant in the {\it same}-hemisphere topology as an unpolarized baseline to the previously extracted azimuthal asymmetries in di-hadron production, related to the spin-dependent {\em interference fragmentation functions} \cite{belleiff}.

The $z$ range from 0.2 to 1.0 is separated into 16 equidistant bins, while the invariant mass is split into 100 uniform bins between 0.3 GeV and 2.3 GeV in order to be able to see the mass structure of the cross section. As $z$ is related to the total energy of the hadron pair, not all masses are necessarily available in a given $z$ bin. All hadron and charge combinations are treated independently in order to test their consistency where applicable (i.e., for charge conjugate states, such as $\pi^+\pi^+$ and $\pi^-\pi^-$). After confirming their consistency, the final cross sections presented here do combine those sets of equal information, leaving 6 combinations in total. 

\subsection{PID correction}

As in Ref.~\cite{martin}, the particle misidentification is corrected using inverted $5 \times 5$ particle-misidentification matrices for the five particle hypotheses (pions, kaons, protons, muons, and electrons) for each identified particle, laboratory momentum, and polar angle bin. These matrices are obtained using decays of $D^{*+}$, $\Lambda$, and $J/\psi$ from data where the true particle type is determined by the charge reconstruction and the invariant mass distribution. Occasionally, when too few events are available in the data, the extracted efficiencies are interpolated and/or extrapolated based on the behavior in the generic MC; this occurs particularly at the boundaries of the acceptance.
The matrices are calculated for each of the two-dimensional bins in laboratory momentum and polar angle, with the boundaries of the $17$ bins in momentum at ($0.5, 0.65, 0.8, 1.0, 1.2, ...., 3.0, 3.5, 4.0, 5.0, 8.0$) GeV and the boundaries of the 9 bins in $\cos\theta_{\mathrm{lab}}$ at ($-0.511, -0.3, -0.152, 0.017, 0.209, 0.355, 0.435, 0.541,$ $ 0.692, 0.842$). 

In this analysis, the inverted misidentification matrix is applied for each of the identified hadrons by multiplying the respective weights for each hadron being a pion or kaon to obtain the total weight for the di-hadron in any of the four pion-kaon combinations. To confirm the consistency of this treatment, the $D^0$ branching ratios for the pion-pion and kaon-kaon decay channels to the pion-kaon decay channel are compared to the PDG \cite{pdg} values and found to be consistent. We also confirm that the total yield of particle pairs is unaffected by this treatment. In particular, for the particle combinations of interest here, there are $2.3\cdot 10^8$ pion pair candidates before PID correction and $2.6\cdot 10^8$ after correction. Similarly, $1.2\cdot 10^8$ pion-kaon pair candidates become $1.1\cdot 10^8$, and $2.3\cdot 10^7$ kaon pair candidates become $2.2\cdot 10^7$ while the sum of all pairs before and after is unchanged at $4.4\cdot 10^8$. With the exception of pion-proton combinations, all other particle combinations are at least one order of magnitude lower.

The corrected yields are distributed among the ($z,m_{h_1h_2}$) bins according to the corresponding hadron masses: one identified hadron pair appears in several bins with the above-determined weights, depending on the particular hadron combination. 
The ratios relative to the uncorrected hadron assignment are displayed in Fig.~\ref{fig:beforeafterpid}, where one can see that the overall corrections are of the order of 20\% with notable excursions. The dip at masses around 1.6 to 1.8 GeV is caused by $D^0$ decays into pion-kaon pairs where the kaon was mis-identified as pion.
For the uncertainties on the PID correction, the individual inverted matrix elements were varied by their uncertainties and the resulting variation was taken into account. In addition, for the matrix elements that required extrapolation, two different types of extrapolation based on the MC were considered with the variation around the central value assigned as an additional systematic uncertainty. 
The total uncertainties due to the PID correction are at the 10\% level at small values of $z$ and moderate masses, increasing towards the highest available masses and decreasing for intermediate $z$ bins. 

\begin{figure*}[th]
\begin{center}
\includegraphics[width=0.95\textwidth]{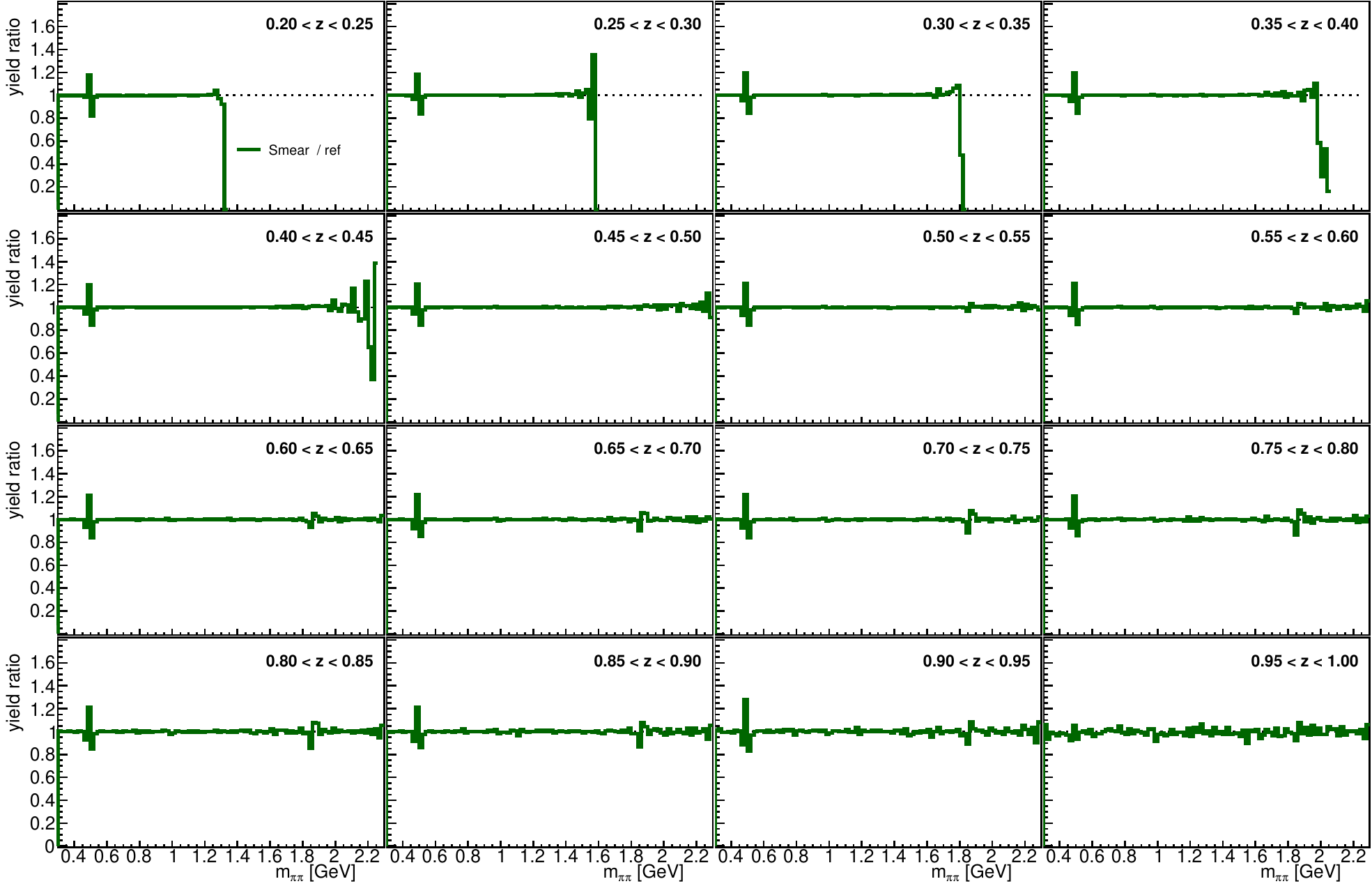} 
\caption{\label{fig:beforeaftersmearing}Ratio of yields after to before applying the smearing correction for $\pi^+\pi^-$ pairs as a function $m_{\pi\pi}$ in bins of $z$. Structures are observed due to sharp resonances; at low $z$, the kinematical limits are evident. }
\end{center}
\end{figure*}

\subsection{Momentum smearing correction}
The momentum smearing is corrected using the singular value decomposition unfolding \cite{hoecker} as implemented in Root \cite{root}. Since not all $z \times m_{h_1h_2}$ bin combinations are kinematically available, a reduced smearing matrix of only the bin combinations with nonzero entries is extracted based on the generic MC (using the true PID) while other MC settings were used as a consistency check. In the remaining bin combinations, the smearing is relatively moderate and the optimal regularization parameter, as prescribed in Ref.~\cite{hoecker}, is generally very close to the full rank of the reduced matrices. This indicates that the statistics of matrix and data vectors is sufficiently large so that fluctuations in the MC statistics do not play a significant role. Generally, the unfolded yields are very similar to the raw ones with only some corrections around very narrow resonances. The final before/after ratio plots are displayed in Fig.~\ref{fig:beforeaftersmearing} as a function of ($z$, $m_{h_1h_2}$) for pion pairs, where one can see that the ratios are predominantly around unity.

All uncertainties prior to the smearing-unfolding (PID and statistical uncertainties) are unfolded as well, resulting in the respective covariance matrices. The covariance matrix due to the MC statistics itself and the differences with an analytic unfolding (i.e., application of the inverted response matrix) are assigned as systematic uncertainties related to the unfolding. These systematic contributions are comparable to the statistical uncertainties and stay below the percent level except for the high-mass tails.

\subsection{Non-$q \bar{q}$ background correction}
\begin{figure*}[ht]
\begin{center}
\includegraphics[width=0.95\textwidth]{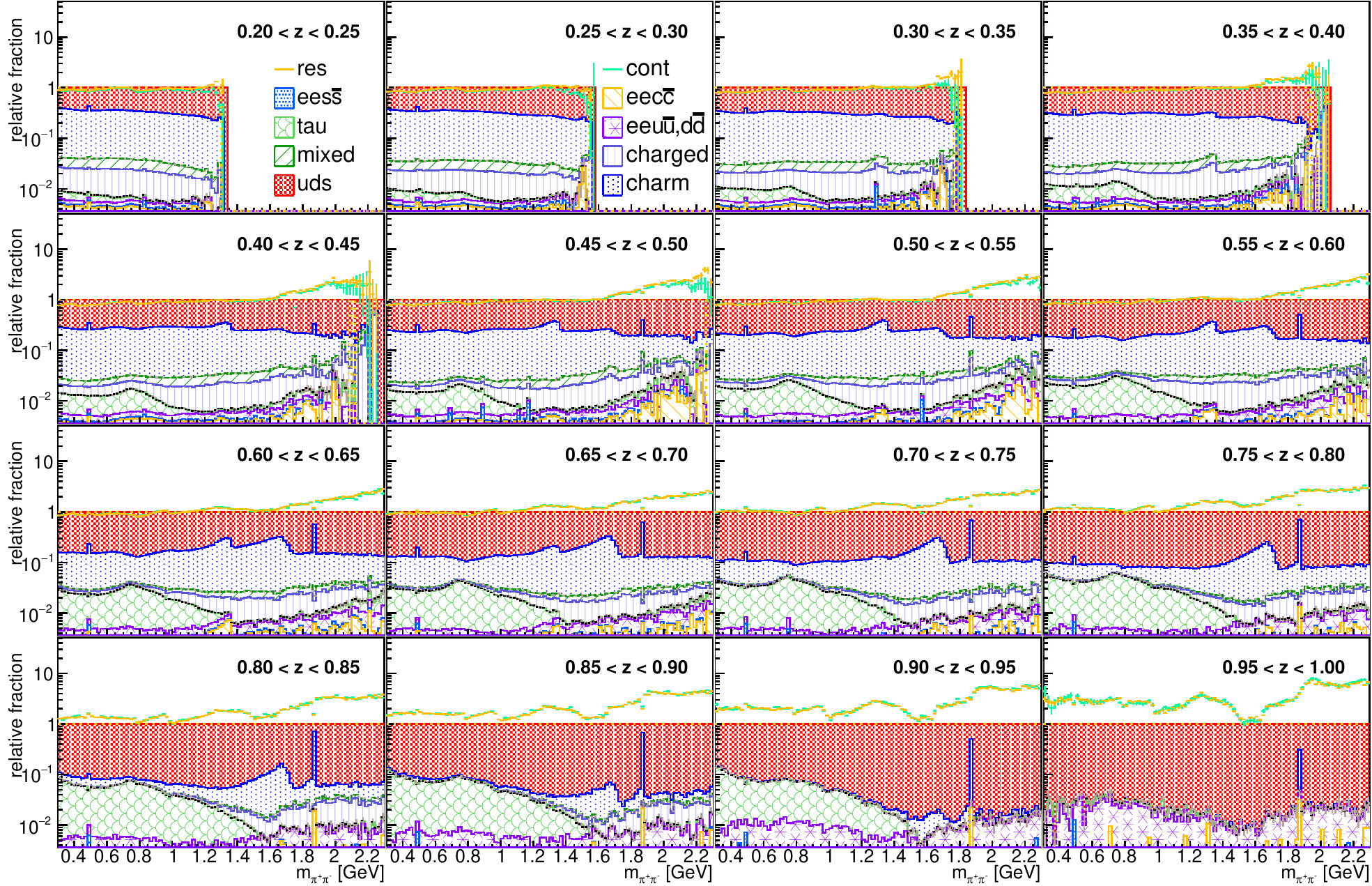}
\caption{\label{fig:zqedcontpid0_mix4}Fraction of $\pi^+\pi^-$ pairs as a function of $m_{\pi\pi}$ in bins of $z$ originating from various sub-processes. The individual relative contributions are displayed from top to bottom for $uds$ (red filled area), charm (blue dotted area), mixed [$\Upsilon (4S)\rightarrow B^0\overline{B}^0$, dark-green hatched area] and charged [$\Upsilon (4S)\rightarrow B^+B^-$, violet horizontally hatched area], $\tau$ pair (light-green scaled area), $eeu\bar{u}+eed\bar{d}$ (purple starred area), $ees\bar{s}$ (light-blue dotted area) and $eec\bar{c}$ (orange hatched area) events. 
Also, for comparison, the continuum (green solid lines) and on-resonance (orange dotted lines) data relative to the MC sum are shown.}
\end{center}
\end{figure*}

Several processes that are not part of the fragmentation-function definition need to be removed from the initial yields. These include the two-photon processes $e^+e^-\rightarrow e^+e^-u\bar{u}$, $e^+e^-\rightarrow e^+e^-d\bar{d}$, $e^+e^-\rightarrow e^+e^-s\bar{s}$ and $e^+e^-\rightarrow e^+e^-c\bar{c}$, as well as $\tau$ pair production and the $\Upsilon(4S)$ decays via either charged or neutral $B$ meson pairs. 
As most of these processes are very well described in simulations, these contributions are directly subtracted from the luminosity-normalized data yields. As can be seen in Fig.~\ref{fig:zqedcontpid0_mix4} for $\pi^+\pi^-$ pairs, the relative contributions are generally small and do not contribute more than a few percent. For pion-kaon and kaon-kaon combinations (not shown here), the $\Upsilon$ decay contributions increase to the 10\% level while $\tau$ pair contributions, which dominate the pion-pair background, are essentially negligible.

For the systematic uncertainties due to the non-$q\bar{q}$ process removal, the statistical uncertainties of the MC samples are taken into account. Also, the contributions are varied by a factor of 1.4\% for the $\tau$ process and a factor of five for the two-photon processes in order to reflect the level of confidence in these simulations \cite{uehara}. Due to the small relative contributions, these systematic uncertainties are generally well below the percent level for pion pairs and at most a few percent for other hadron combinations.  
\subsection{Preselection and acceptance correction}
Another correction treats the reconstruction efficiencies due to particle selection and tracking efficiencies as well as acceptance effects. 
\subsubsection{Reconstruction efficiency within the barrel acceptance}
\begin{figure*}[ht]
\begin{center}
\includegraphics[width=0.95\textwidth]{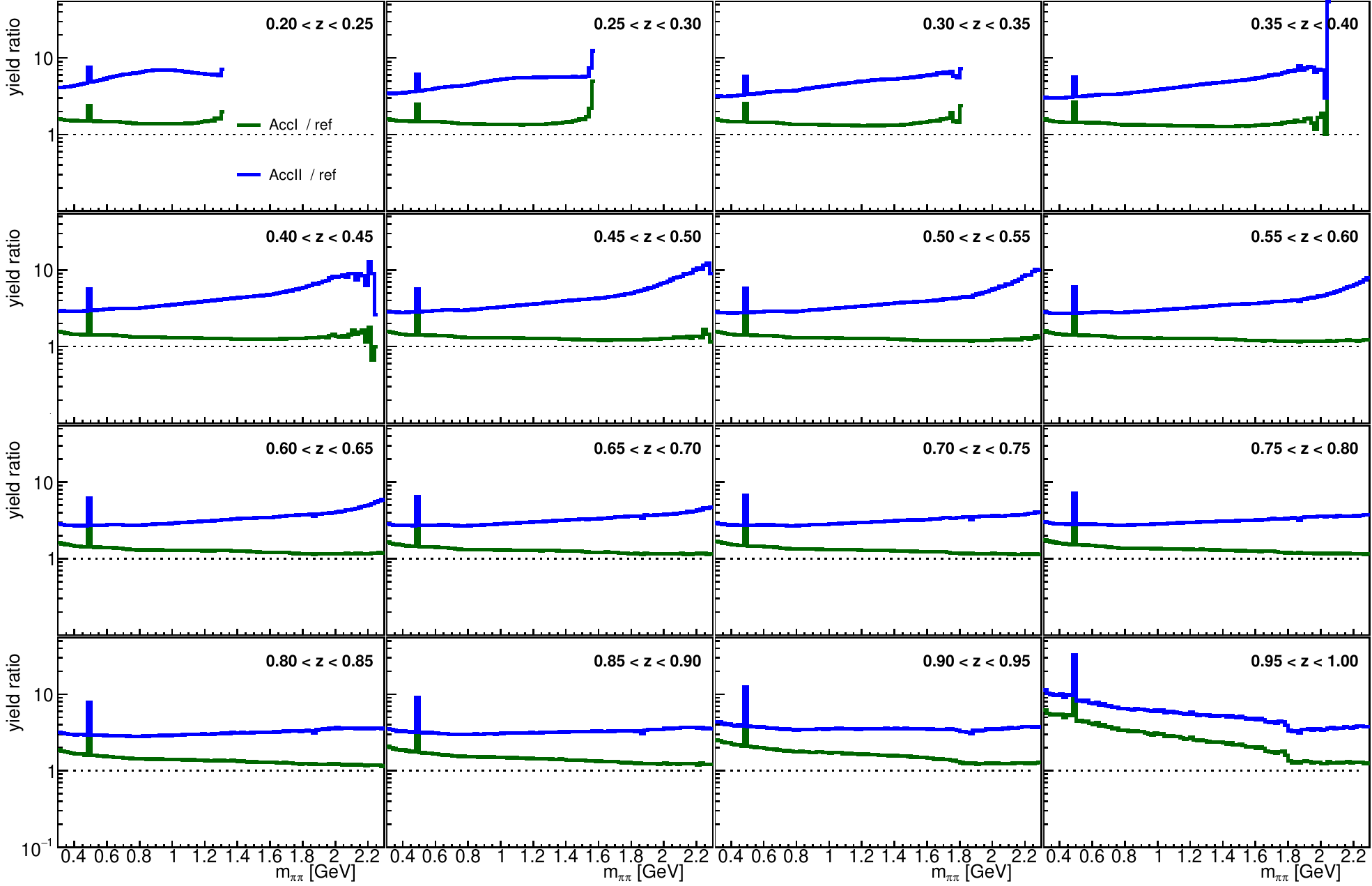} 
\caption{\label{fig:acceptance}Yield ratios after successively applying all acceptance and efficiency corrections (labeled AccI and AccII as the two acceptance corrections discussed in the text) relative to the reference yields after non-$q\bar{q}$ removal for $\pi^+\pi^-$ pairs as a function of $m_{\pi\pi}$ in bins of $z$.}
\end{center}
\end{figure*}

The first of these corrections takes into account the event preselection as well as particle reconstruction efficiencies due to the various selection criteria. It is calculated by building the ratios of yields between reconstructed and generated hadron pairs using the correct momenta and PID. However, the thrust direction cut, as well as the minimum momentum of the individual hadrons and $z$ requirements, are still applied. The correction factors for $\pi^+\pi^-$ pairs, which are summarized in Fig.~\ref{fig:acceptance}, show that the corrections are relatively flat and smooth as a function of both the invariant mass as well as the fractional energy. The only striking deviation can be found at the $K^0_S$ mass and originates in the SVD hit number requirement. As a $K^0_S$ meson does not decay immediately, especially at higher $z$, the decay pions are formed after the first SVD layers and thus the SVD hit requirement cannot be fulfilled. With this exception, the reconstruction efficiencies, which inversely affect the correction factor, range between 20\% at very high $z$ and low masses to about 80\% at higher masses and intermediate $z$.
The behavior for pion-kaon and kaon pairs is similar (not shown here) but, lacking substantial contributions from long-lived resonances, no substructure in the correction factors as a function of mass or $z$ is visible. These reconstruction efficiencies are generally smaller than those for pion pairs.

\subsubsection{Acceptance outside the barrel region}
The second correction is evaluated by calculating the ratios between generated hadron-pair yields within the barrel acceptance and those taking the full acceptance into account. The thrust-axis direction is now allowed to be anywhere and the minimum lab momentum requirement necessary for the PID correction is removed. Also, the minimal transverse momentum requirement is here adjusted while the minimal fractional energy requirement for each individual hadron is not. The acceptance functions decrease with increasing mass due to the larger opening angles, while slightly increasing in $z$ due to the larger boosts. The efficiencies, which inversely affect the correction factor, range up to 60\% at the lowest masses and intermediate to high $z$ bins, and drop below 20\% at the highest masses studied. The pion-kaon and kaon-pair efficiencies in addition display a stronger mass dependence, at very small $z$, where the efficiencies drop below 10\% due to the limited accepted phase space, while otherwise having similar magnitudes.\par 
 
The two sequentially applied acceptance corrections are shown in Fig.~\ref{fig:acceptance} for $\pi^+\pi^-$ pairs. For both corrections, the statistical uncertainties in the MC on which these corrections are based are taken as systematic uncertainties. They are comparable to the statistical uncertainties in the data themselves. In addition, the variation of acceptance efficiencies for different fragmentation tunes in the MC is also assigned as a systematic uncertainty. While the responses of the tunes are very similar, the relatively low efficiencies amplify the variations to be generally larger than the statistical uncertainties. Corrections for the same-sign hadron pairs are similar, though generally slightly larger. 

Changes to the partial-wave composition of the di-hadron cross sections due to the acceptance and reconstruction restrictions are not taken into account in this analysis. In general, the MC, when including all corresponding selection criteria as for real data, describes the data well in the measured region, lending support to its use for potential corrections. See Appendix \ref{sec:partialwaves} for a more detailed discussion about the selection effects on the partial-wave composition.

\subsection{Weak decays}
\begin{figure*}[ht]
\begin{center}
\includegraphics[width=0.95\textwidth]{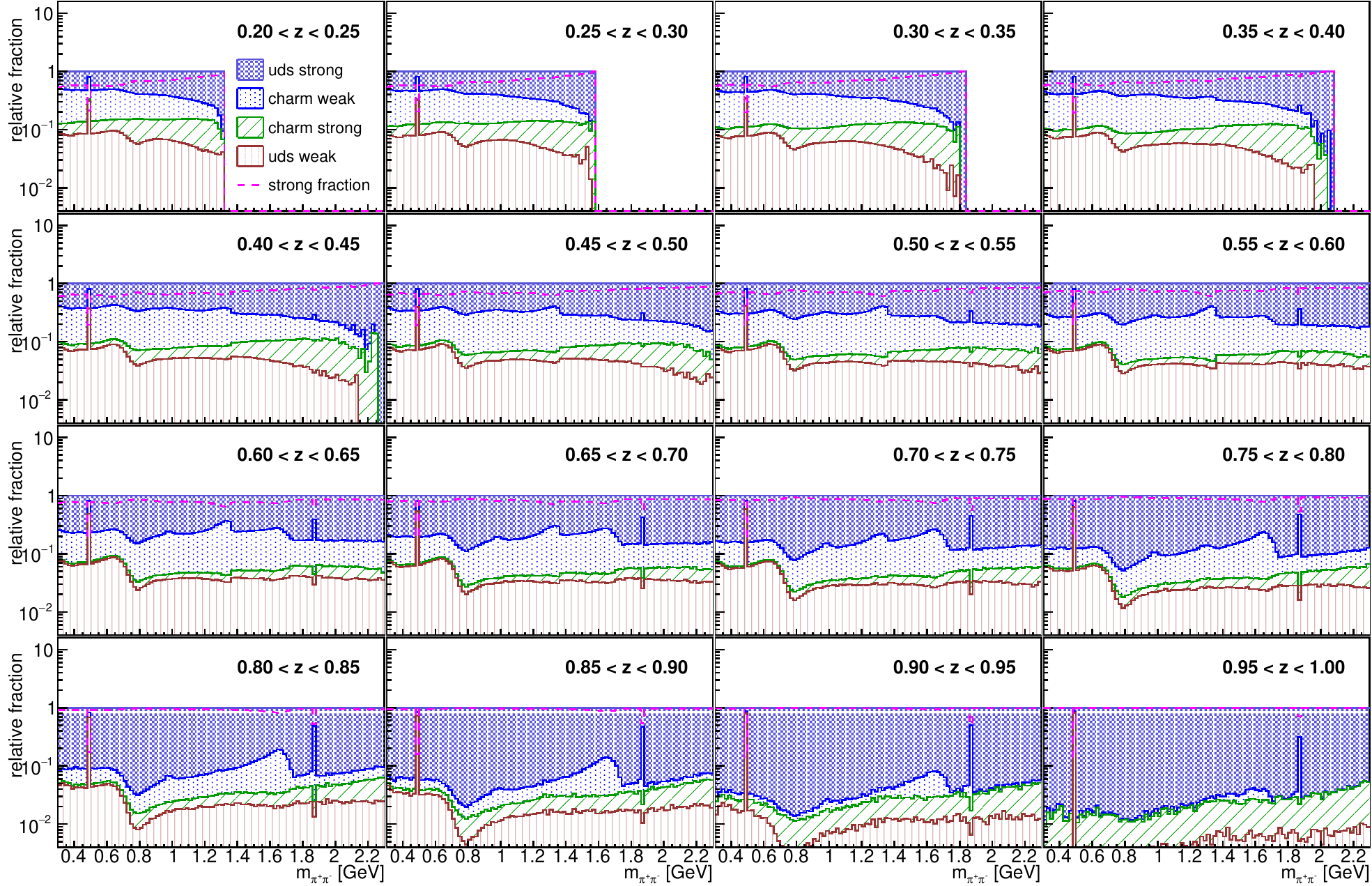}

\caption{\label{fig:zweakratio_mix0} Fraction of $\pi^+\pi^-$ pairs as a function of $m_{\pi\pi}$ in bins of $z$ originating from weak and strong decays. The individual relative contributions are displayed from top to bottom for strong $uds$ decays (purple dark-filled area), weak charm decays (blue dotted area), strong charm decays (dark-green negative-hatched area) and weak $uds$ decays (red horizontally striped area). The strong decay fractions are also displayed as dashed magenta lines.}
\end{center}
\end{figure*}

The definition of fragmentation functions, and in particular the application of QCD-only DGLAP equations \cite{dglap}, in principle requires only products of strong processes and decays to be taken into account. This is experimentally not possible and either relies on partial removal of experimentally accessible decays or full removal based on MC simulations. As many published results do not remove weak decays at all, we will provide both results before and after subtraction of all weak decays based on the MC. For the latter, ancestries of each hadron were traced back to either a weak decay or the initial gluonic strings. The relative contributions of weak and strong decays can be seen in Fig.~\ref{fig:zweakratio_mix0} for $\pi^+\pi^-$ pairs. As expected, most charm events include weak decays in order to arrive at pions or kaons while $uds$ events are generally dominated by strong decays. The overall fraction of strong decay di-pions stays above 50\% and increases to about 90\% at higher $z$, while it dips to below 30\% at the $K^0_S$ and $D^0$ masses. For pion-kaon and kaon pairs, the strong-decay fractions are generally lower due to the larger charm contributions in these samples and the preferred decay of the various charm contributions into kaons.

\subsection{ISR correction}
\begin{figure*}[ht]
\begin{center}
\includegraphics[width=0.95\textwidth]{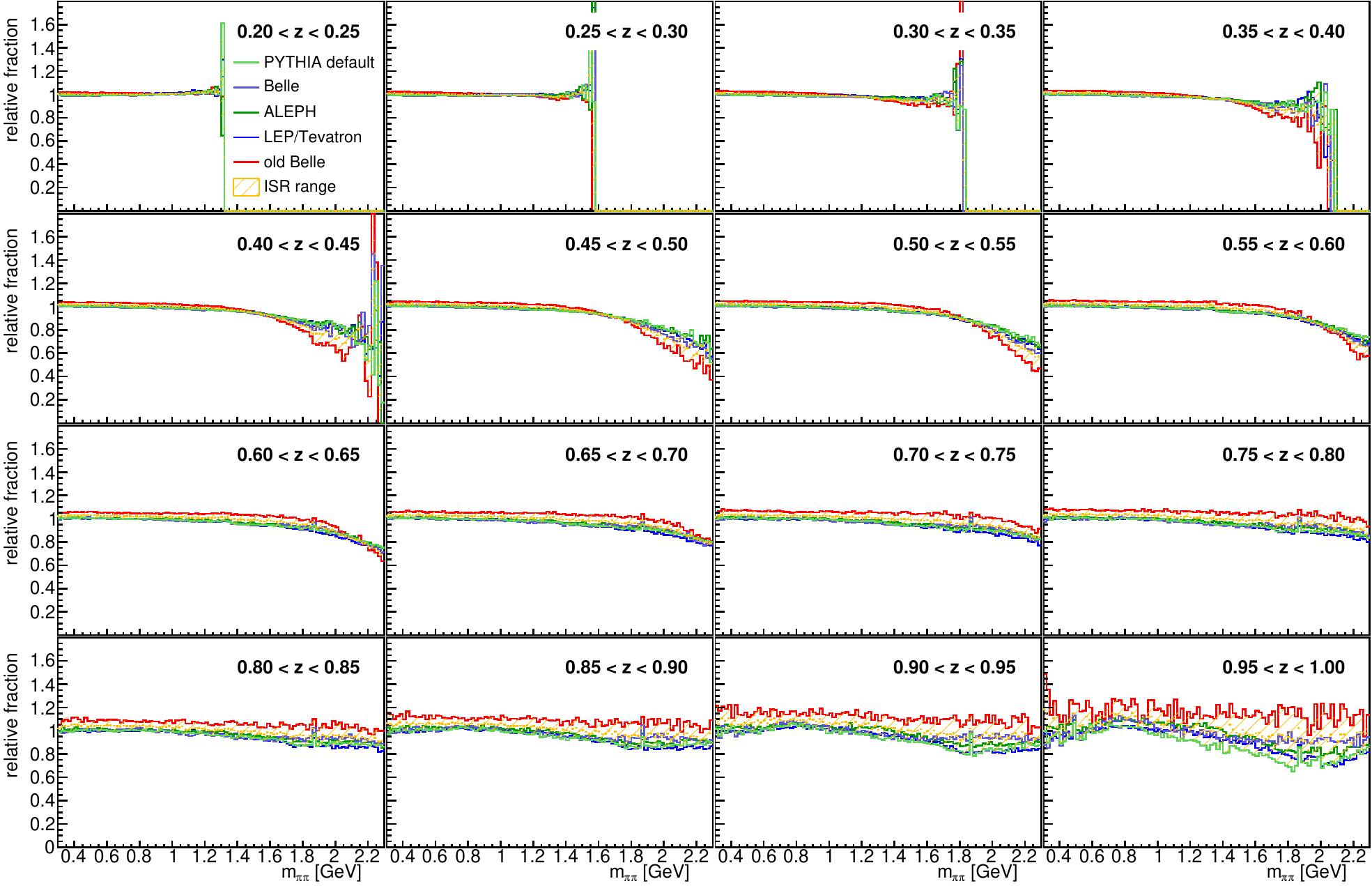}
\caption{\label{fig:isrzfraction_pid0_mix0}Non-ISR over ISR ratios of $\pi^+\pi^-$ pairs as a function of $m_{\pi\pi}$ in bins of $z$.}
\end{center}
\end{figure*}

The last correction is for the initial-state radiation (ISR) effects. Unlike the previous publications \cite{martin,dihadronprd1}, a more rigorous correction is applied. The effect of initial-state radiation is studied by comparing the generated MC cross sections with ``no ISR'' and ``including ISR'' by application of the {\sc Pythia} switch MSTP(11). The ratios between these options can be seen in Fig.~\ref{fig:isrzfraction_pid0_mix0} for $\pi^+\pi^-$ pairs. At small masses, the $z$ behavior is as expected. At low $z$, ISR yields are slightly larger than the non-ISR yields due to ISR events being able to feed-down to low-$z$. The CMS energies are no longer sufficient with ISR to populate the higher-$z$ regions and thus the non-ISR cross sections are larger in these regions. At larger masses, this behavior changes drastically and the cross sections including ISR become more than 30\% larger than those for non-ISR events. In this case, it is found that when a substantial amount of energy is taken away via ISR photons, the nominal boost, and consequently the hemisphere definitions, become incorrect and ISR events from opposite ``true'' hemispheres enter the yields. This has been verified by explicitly calculating the true hemispheres of hadron pairs in the additionally boosted quark-antiquark system in the presence of ISR photons.  
These non-ISR to ISR ratios are then used to correct the data using the {\sc Pythia} default settings. The variations between different MC tunes were taken into account as systematic uncertainties. While the behavior is nearly identical in most settings, the corrections are substantially different for the generic Belle simulations and thus the total systematic uncertainties are dominated by this contribution. For all hadron combinations, the uncertainties range between several percent at low $z$ and masses to close to 100\% at high masses.   

\subsection{Consistency checks and total systematic uncertainties\label{sec:syst}}

After all the corrections are applied, we perform several consistency tests. With the removal of the $\Upsilon(4S)$ decays, the data at the $\Upsilon(4S)$ resonance as well as the data from 60 MeV below are found to be consistent, as they should contain the same information. Also, the opposite charge-sign combinations are found to be consistent and can be merged where applicable. At the end of these and similar checks, the final di-hadron cross sections and their statistical and total systematic uncertainties can be evaluated for the 6 unique di-hadron combinations.
The total systematic uncertainties for the cross sections were conservatively taken as the linear sum of the contributing hadron-pair uncertainties as calculated in the respective independent analyses for different charge combinations.

\begin{figure*}[htb]
\begin{center}
\includegraphics[width=0.95\textwidth]{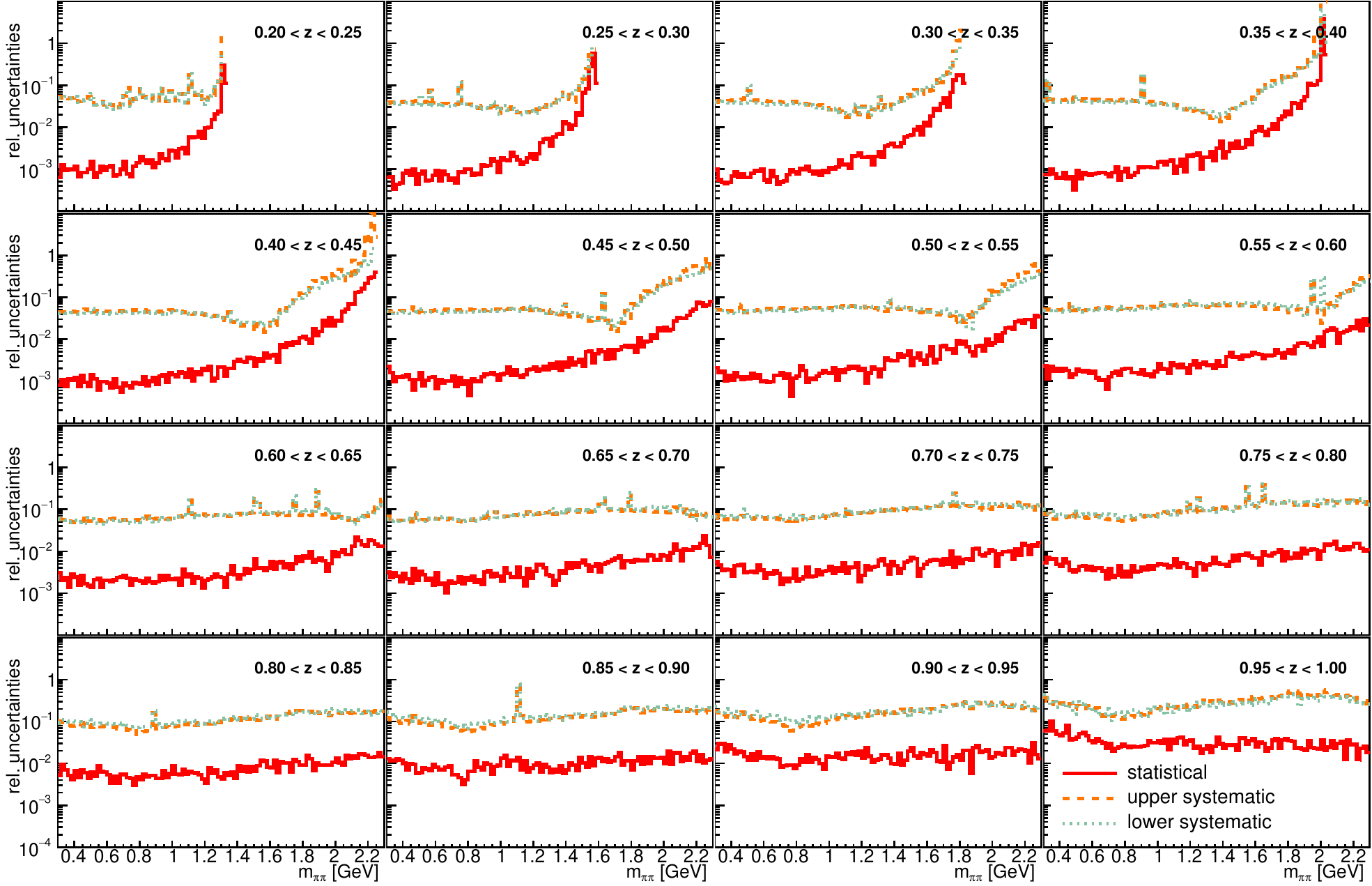}
\caption{\label{fig:systall_mix0} Relative (asymmetric) systematic uncertainties (upper, dashed lines; lower, dotted lines) and statistical uncertainties (full lines) for opposite-sign pion pairs as a function of $m_{\pi\pi}$ in bins of $z$.}
\end{center}
\end{figure*}

Figure \ref{fig:systall_mix0} shows the statistical and total systematic uncertainties for opposite-sign pion pairs. The individual systematic uncertainties from the various correction stages are added in quadrature. The resulting uncertainties are all dominated by the systematics, which are in turn dominated by the ISR systematics at higher $z$ and the PID systematics at lower $z$, especially for kaon combinations, as well as uncertainties due to the acceptance correction. At lower $z$ and lower masses, the total pion-pair systematic uncertainties are below 10\%, while at higher masses, both uncertainties can reach more than 100\%. The behavior for pion-kaon and kaon pairs as well as the same-sign combinations is generally similar and also dominated by systematic uncertainties. 

Some spikes in the systematic uncertainties occur as a result of some large uncertainties in the PID correction matrices from rare, off-diagonal entries. 
Additionally, there are global scale uncertainties due to the luminosity measurement (1.4\%) and the track reconstruction (2$\times$0.35\%) that are not shown, leading to an overall 1.6\% scale uncertainty.  

\section{Results\label{sec3}}
\begin{figure*}[htb]
\begin{center}
\includegraphics[width=0.95\textwidth]{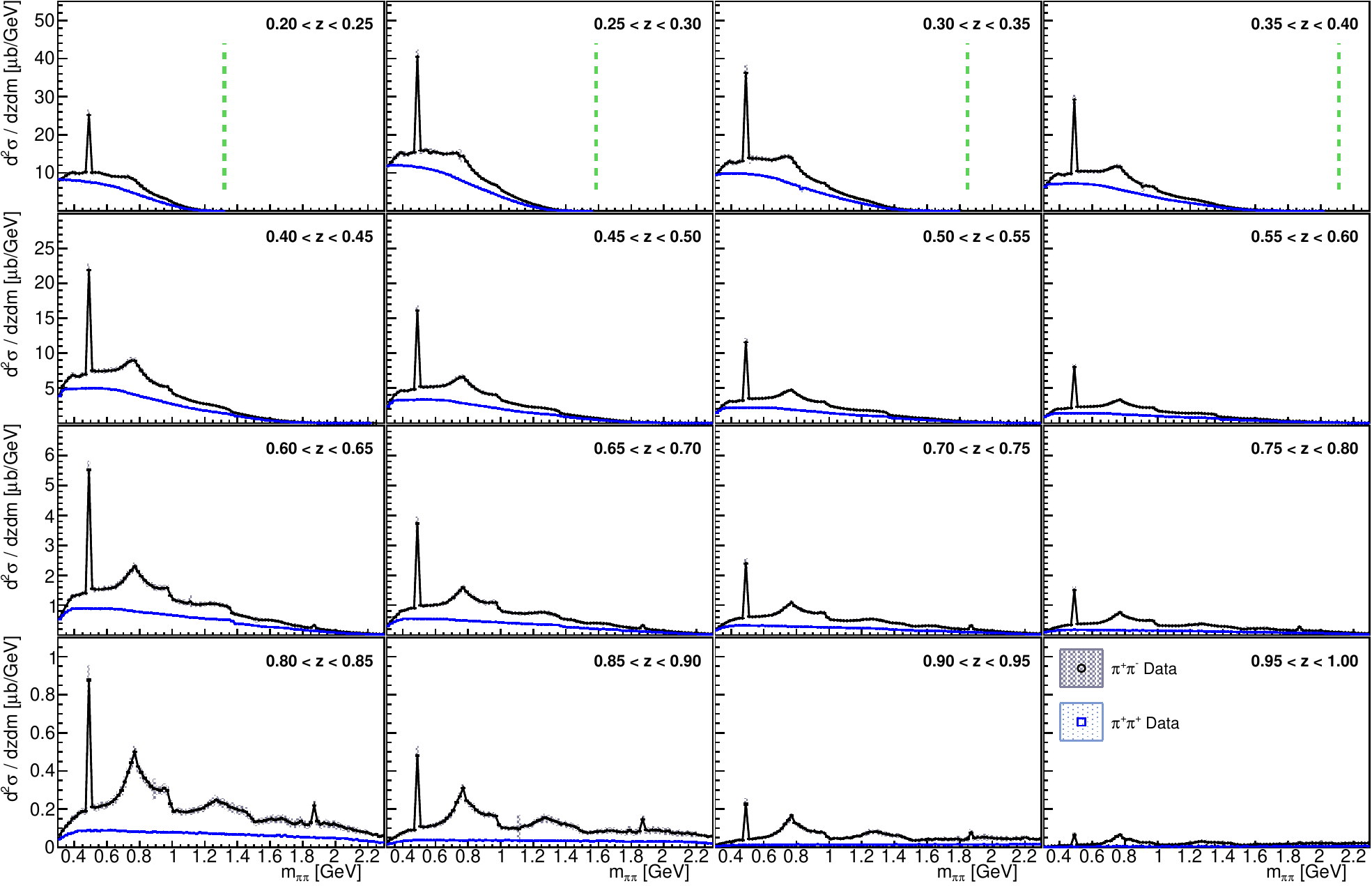}
\includegraphics[width=0.95\textwidth]{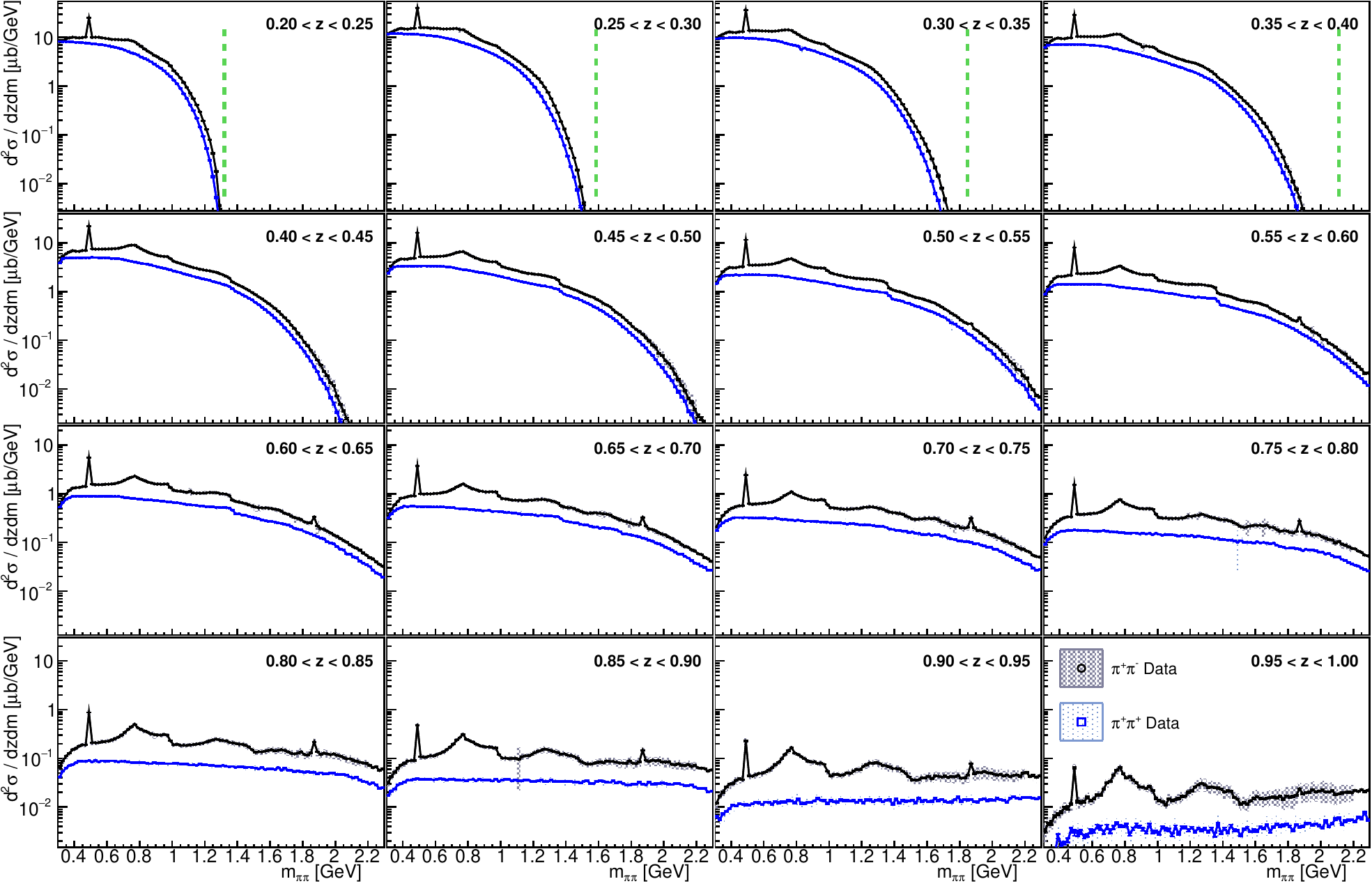}

\caption{\label{fig:allxsec_mix1_pid3}Differential cross sections for $\pi^+\pi^-$ (black circles) and $\pi^+\pi^++\ c.c.$ (blue squares) as a function of $m_{\pi\pi}$ for the indicated $z$ bins. The error boxes represent the systematic uncertainties. Top panel: linear representation of cross sections: bottom panel: logarithmic representation. The vertical green dashed line corresponds to the kinematic limit. An overall 1.6\% scale uncertainty is not shown.}
\end{center}
\end{figure*}

The final di-hadron cross sections for a minimum thrust of 0.8 and minimum individual fractional energies $z_{1,2}$ of 0.1 are displayed in Fig.~\ref{fig:allxsec_mix1_pid3} for pion pairs as a function of $m_{\pi\pi}$ in bins of $z$. The very prominent resonances seen are the $K^0_S$, $\rho^0$ and the Cabibbo-suppressed $D^0$ decays. Based on MC simulations, the enhancements at around 1.35 GeV and 1.6 GeV can be identified as multi-body or indirect decay products of $D$ mesons as well. At around 1 GeV, the $f_0(980)$ can be seen. Part of the cross section at low masses, below the $\rho^0$ resonance, originates from partially reconstructed $\omega$ and $\eta$ decays. The same-sign pion pairs generally display a continuous distribution with a slight enhancement at around 1.35 GeV, which again is caused by decay products of $D$ mesons. 

\begin{figure*}[htb]
\begin{center}
\includegraphics[width=0.95\textwidth]{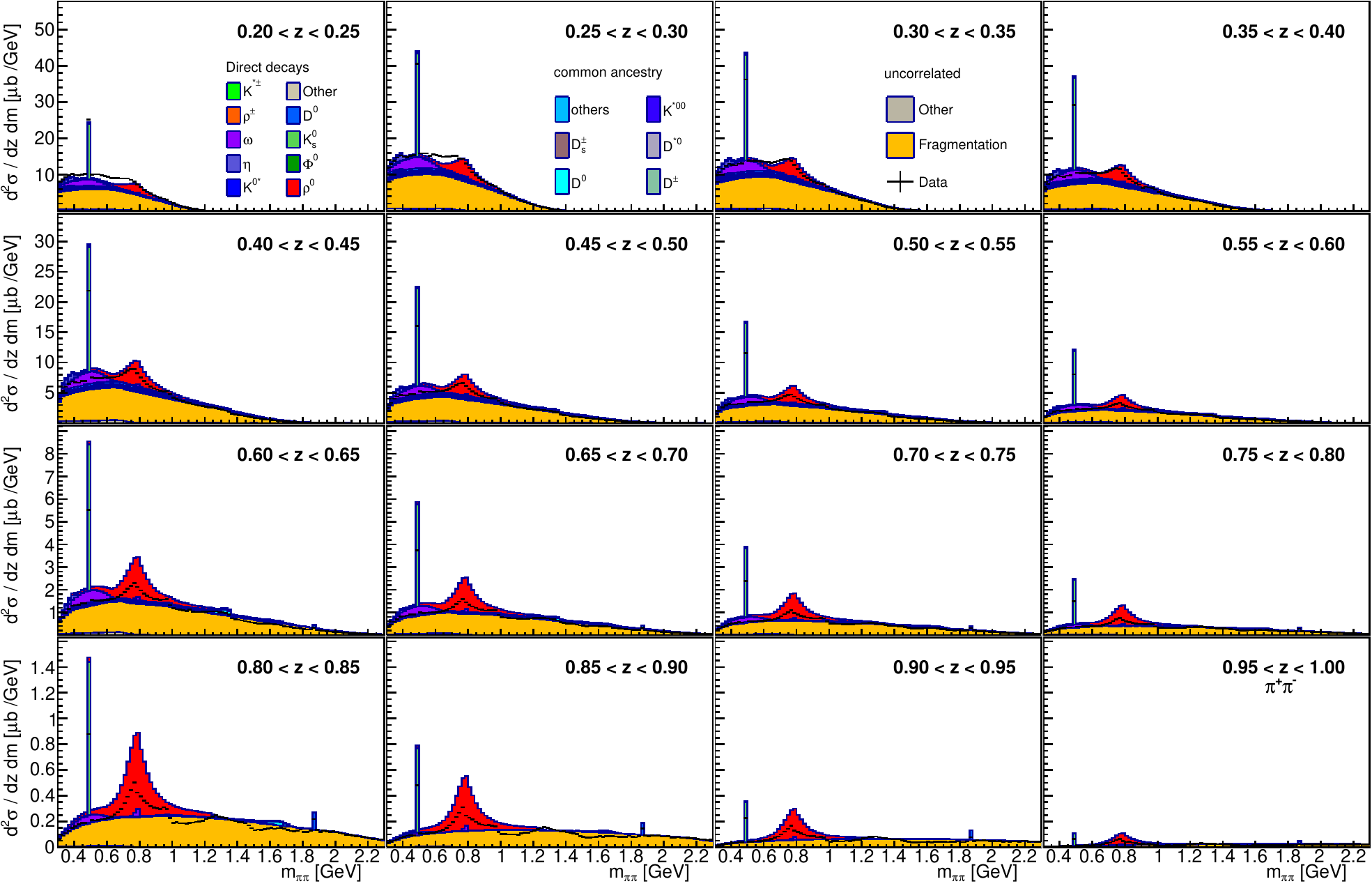}
\includegraphics[width=0.95\textwidth]{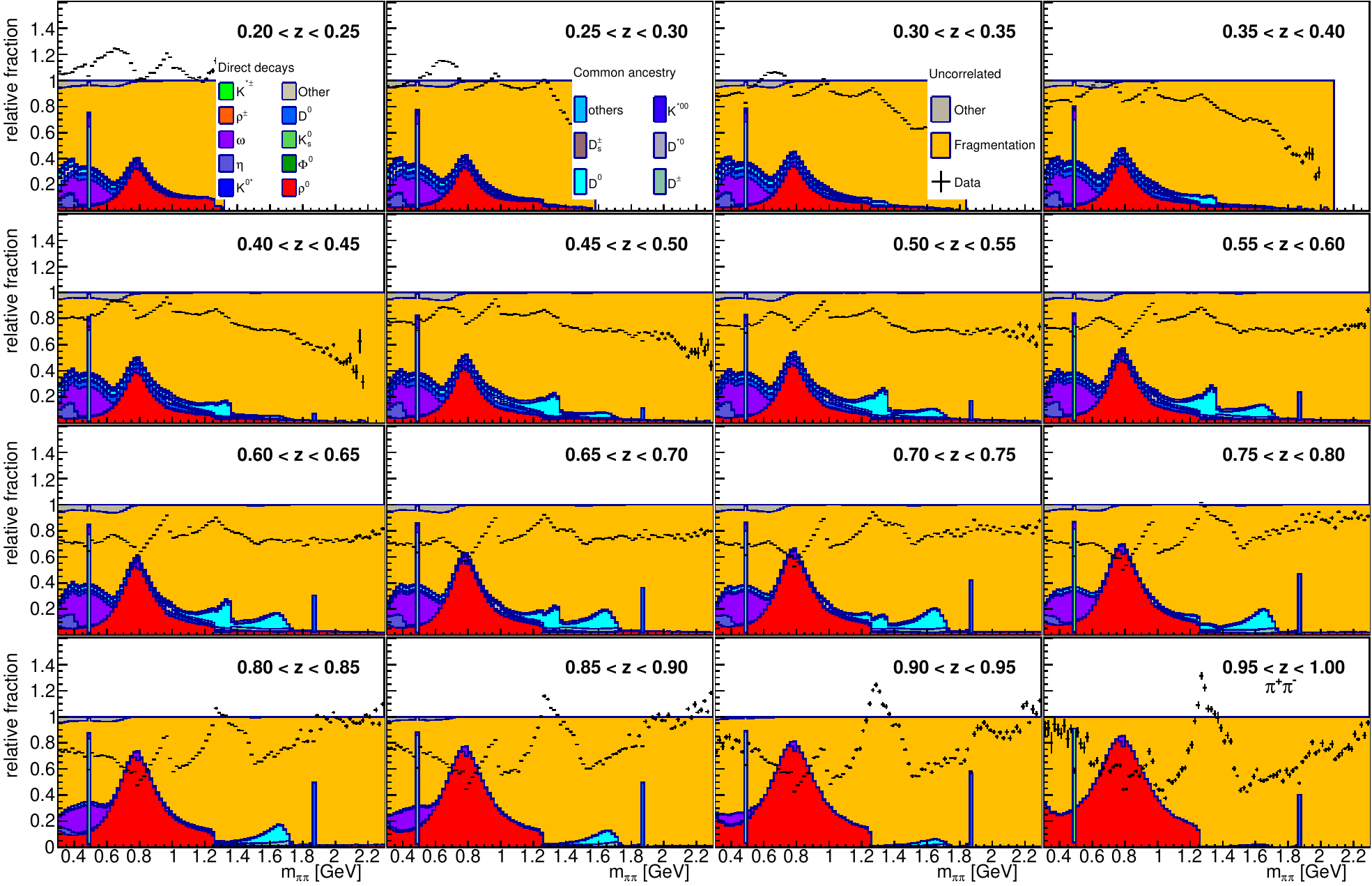}

\caption{\label{fig:results:massdists0} MC decomposition of the unlike-sign pion pairs as a function of $m_{\pi\pi}$ in bins of $z$ for various resonance, partial resonant and non-resonant parents, displayed in linear scale (top) and as a relative fraction of the total cross section (bottom).}
\end{center}
\end{figure*}

The origins of the di-pions we observe is in many cases only accessible via MC and it is informative to discuss them further. In Fig.~\ref{fig:results:massdists0}, the stacked absolute and relative contributions are displayed for unlike-sign pion pairs, separated by parentage of both pions according to their common ancestor. One clearly sees several of the direct two- or three-body decays such as $K_S^0$, $\rho^0$, $D^0$, $\eta$, etc., as well as pions from different steps in $D$ meson decay chains. A substantial fraction of pion pairs have no common ancestor: they originate directly from the fragmentation chain. 
Note that the abrupt drop of the $\rho^0$ contribution at around 1.25 GeV is an artifact of the MC generator itself and in reality the Breit-Wigner shape is expected to extend to higher masses. 

\begin{figure*}[htb]
\begin{center}
\includegraphics[width=0.95\textwidth]{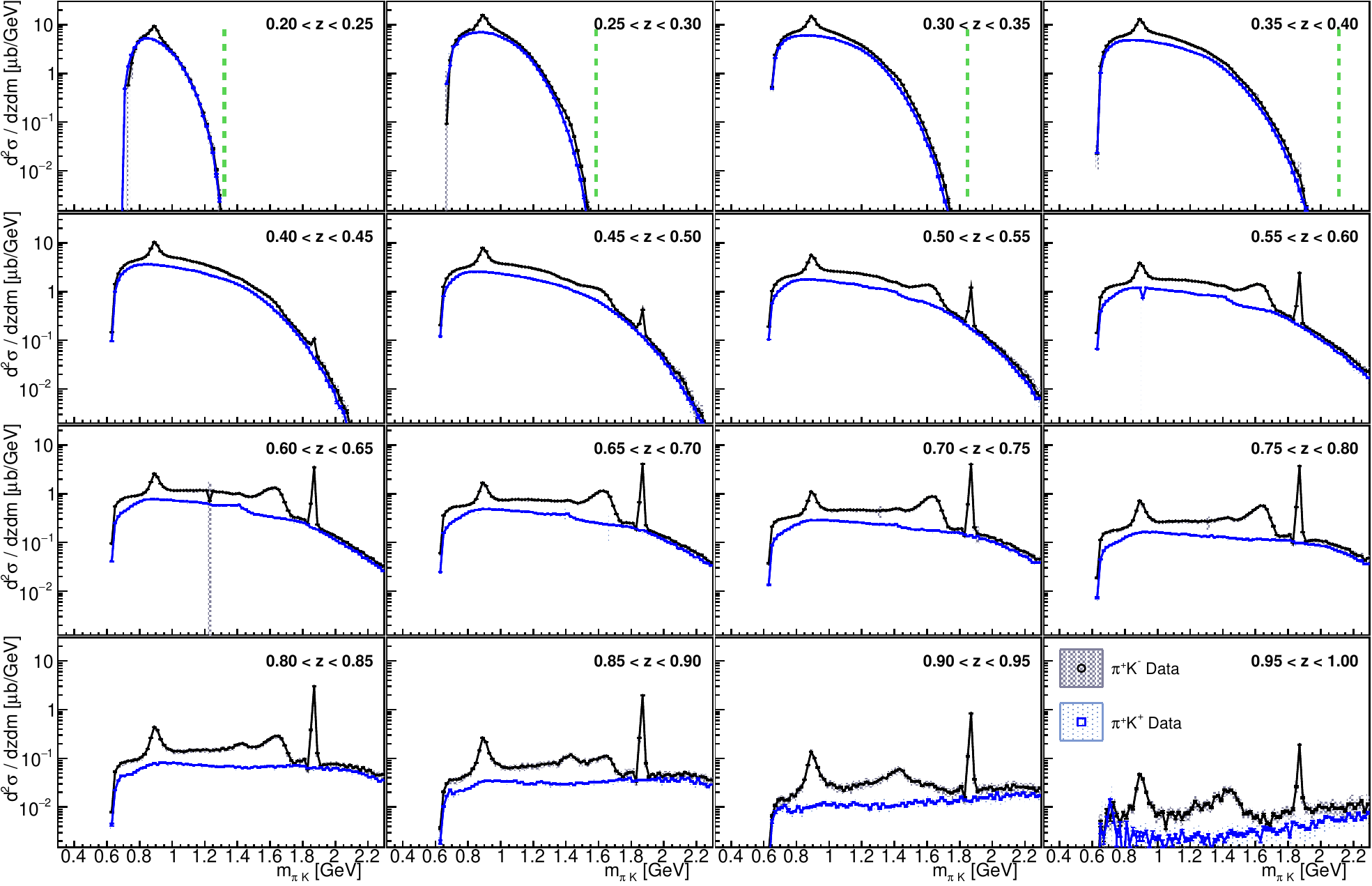}

\caption{\label{fig:allxsec_mix1_pid48}Differential cross sections for $\pi^+K^-+\ c.c.$ (black circles) and $\pi^+K^++\ c.c.$ (blue squares) as a function of $m_{\pi K}$ for the indicated $z$ bins. The error boxes represent the systematic uncertainties. The vertical green dashed line corresponds to the kinematic limit. An overall 1.6\% scale uncertainty is not shown.}
\end{center}
\end{figure*}

The pion-kaon cross sections as a function of invariant mass and $z$ bin can be seen in Fig.~\ref{fig:allxsec_mix1_pid48}. The corresponding $K^*$ resonance and the Cabibbo-favored $D^0$ meson decay are very clearly visible. The enhancement at 1.6 GeV is again predominantly caused by $D$ meson decays. No further resonances are easily identifiable. The same-sign pion-kaon pairs again show a predominately smooth distribution from direct fragmentation.

\begin{figure*}[hbt]
\begin{center}
\includegraphics[width=0.95\textwidth]{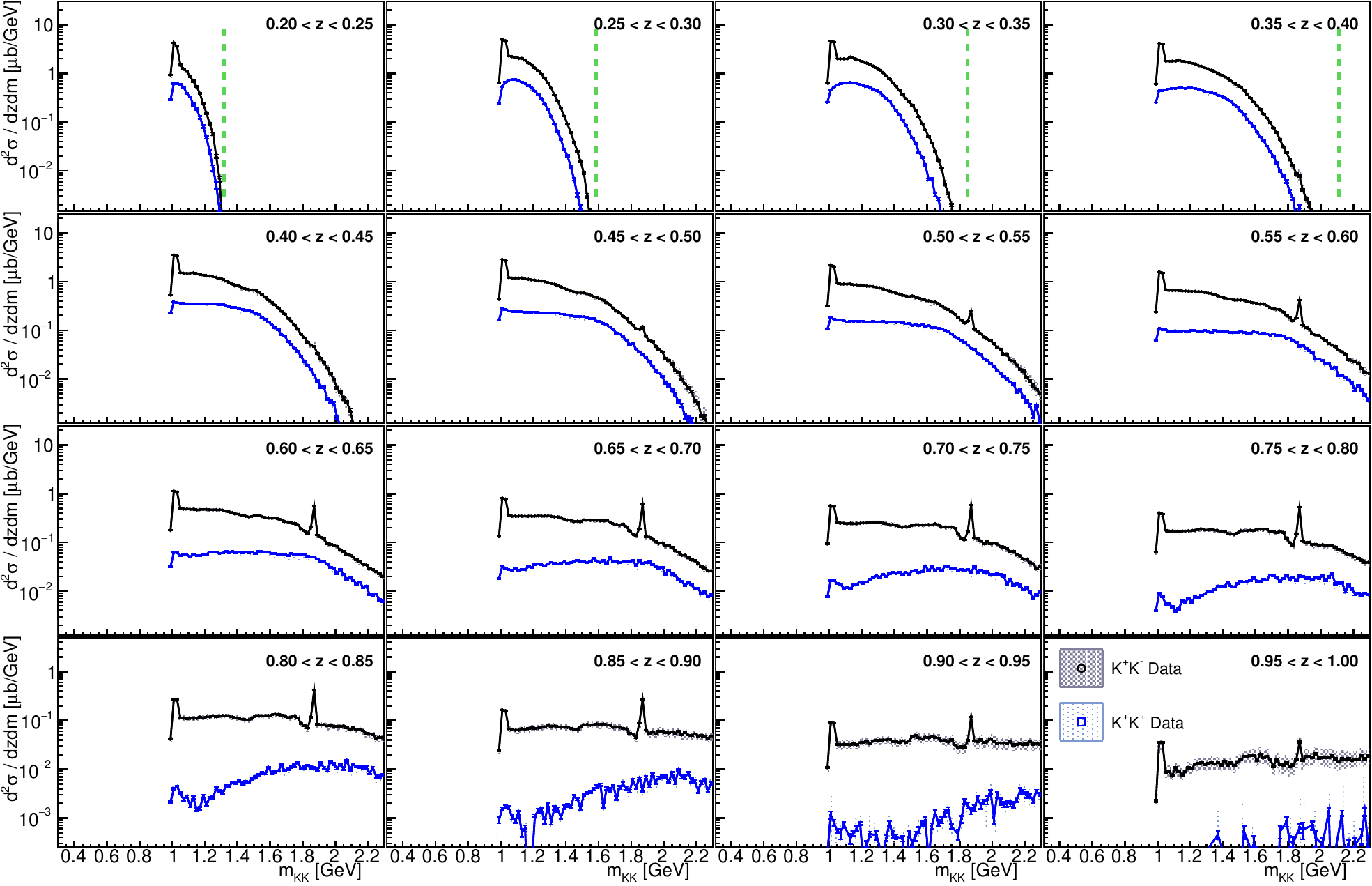}

\caption{\label{fig:allxsec_mix1_pid12288}Differential cross sections for $K^+K^-$ (black circles) and $K^+K^++\ c.c.$ (blue squares) as a function of $m_{KK}$ for the indicated $z$ bins. The error boxes represent the systematic uncertainties. The vertical green dashed line corresponds to the kinematic limit. An overall 1.6\% scale uncertainty is not shown.}
\end{center}
\end{figure*}

Finally, cross sections for kaon-pairs as a function of invariant mass and $z$ bin are displayed in Fig.~\ref{fig:allxsec_mix1_pid12288}. The $\phi$ resonance and (again) the Cabibbo-suppressed $D^0$-meson decay are clearly visible. Some enhancements around the $D^0$ mass can be assigned to $D$ meson decays, except that here $D_s$ mesons seem to play a bigger role consistent with the additional strangeness in the selected hadrons. The same-sign kaon pair cross section is again mostly a smooth function of the invariant mass and is increasingly suppressed with $z$. The small enhancement close to the mass threshold can be, according to MC simulations, potentially related to $D^0$-meson decays into a kaon and another hadron, such as the $a_1$ meson, which can further decay into more kaons. 

The weak-decay-removed cross sections are displayed and discussed in Appendix \ref{sec:weakdecays}. The remaining strong decays do contribute substantially to the total di-hadron cross sections in agreement with model predictions \cite{Matevosyan:2013aka}.

The data tables for the cross sections presented above and the corresponding tables for the weak-decay-removed data are provided online \cite{onlineref}. 

\subsection{MC generator comparison}
\begin{figure*}[htb]
\begin{center}
\includegraphics[width=0.8\textwidth]{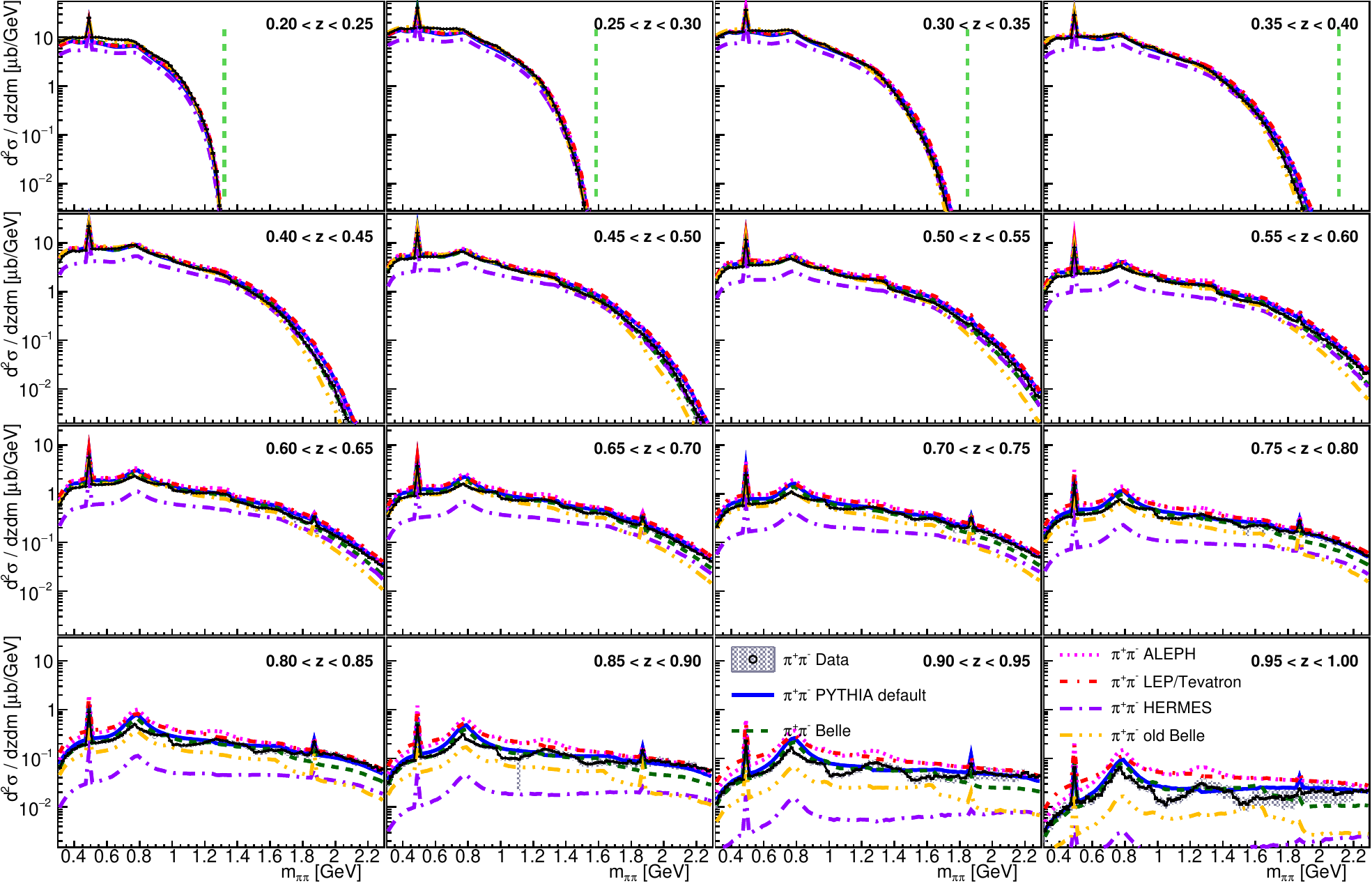}
\caption{\label{fig:allxsec_pid11_mix1_c8_32518153}Differential cross sections for $\pi^+\pi^-$ pairs as a function of $m_{\pi\pi}$ in bins of $z$. Various {\sc Pythia} tunes are also displayed as described in the text. The vertical green dashed line corresponds to the kinematic limit. An overall 1.6\% scale uncertainty is not shown.}
\end{center}
\end{figure*}

Simulations using various {\sc Pythia} settings are displayed in Fig.~\ref{fig:allxsec_pid11_mix1_c8_32518153} and compared to the di-hadron cross sections for the unlike-sign pion pairs. In contrast to the previously published $z_1,z_2$ dependences, no choice describes the cross sections particularly well, while the overall magnitude and $z$ dependence is again best described by the {\sc Pythia} default and the updated Belle simulation settings. Qualitatively, the mass behavior is best described by the ALEPH tune although the magnitude at higher $z$ is too large. The reason for the better mass description is likely the different vector-meson and exited-meson parameter values {\it PARJ(11-17)}, which particularly impact the range between 1.1 GeV and 1.6 GeV. The strength of the $f_0(980)$ seems to be underestimated by all tunes. For pion-kaon and kaon pairs, the behavior is generally similar except that the default settings describe the mass dependence better. This similarity indicates that the differences in the excited meson settings play only a minor role in these comparisons to the data. 

\section{Summary\label{sec4}}
We have reported same-hemisphere di-hadron cross sections as a function of invariant mass and fractional energy for all charged pion and kaon combinations. The measurements will allow a more quantitative application of the previously published polarized di-hadron asymmetries in extracting quark transversity distributions and their tensor charges from the corresponding polarized SIDIS and proton-proton collision data. 
In addition, the cross sections should help to better constrain the fragmentation function parameters in MC simulations that are relevant to studies of either nucleon structure or the size of backgrounds in flavor physics. Apart from a few distinct resonances, the whole mass spectrum has not been measured before.

\begin{acknowledgments}
  We thank the KEKB group for the excellent operation of the accelerator; the KEK cryogenics group for the efficient operation of the solenoid; and the KEK computer group, the National Institute of Informatics, and the PNNL/EMSL computing group for valuable computing and SINET5 network support.  We acknowledge support from the Ministry of Education, Culture, Sports, Science, and Technology (MEXT) of Japan, the Japan Society for the Promotion of Science (JSPS), and the Tau-Lepton Physics Research Center of Nagoya University; the Australian Research Council; Austrian Science Fund under Grant No.~P 26794-N20; the National Natural Science Foundation of China under Contracts No.~10575109, No.~10775142, No.~10875115, No.~11175187, No.~11475187, No.~11521505 and No.~11575017; the Chinese Academy of Science Center for Excellence in Particle Physics; the Ministry of Education, Youth and Sports of the Czech Republic under Contract No.~LTT17020; the Carl Zeiss Foundation, the Deutsche Forschungsgemeinschaft, the Excellence Cluster Universe, and the VolkswagenStiftung; the Department of Science and Technology of India; the Istituto Nazionale di Fisica Nucleare of Italy; the WCU program of the Ministry of Education, National Research Foundation (NRF) of Korea Grants No.~2011-0029457, No.~2012-0008143, No.~2014R1A2A2A01005286, No.~2014R1A2A2A01002734, No.~2015R1A2A2A010032\\\noindent 80, No.~2015H1A2A1033649, No.~2016R1D1A1B0101013\\\noindent 5, No.~2016K1A3A7A09005603, No.~2016K1A3A7A0900\\\noindent 5604, No.~2016R1D1A1B02012900, No.~2016K1A3A7A0\\\noindent 9005606, No.~NRF-2013K1A3A7A06056592; the Brain Korea 21-Plus program, Radiation Science Research Institute, Foreign Large-size Research Facility Application Supporting project and the Global Science Experimental Data Hub Center of the Korea Institute of Science and Technology Information; the Polish Ministry of Science and Higher Education and 
the National Science Center; the Ministry of Education and Science of the Russian Federation and the Russian Foundation for Basic Research; the Slovenian Research Agency; Ikerbasque, Basque Foundation for Science and MINECO (Juan de la Cierva), Spain; the Swiss National Science Foundation; the Ministry of Education and the Ministry of Science and Technology of Taiwan; and the U.S.\ Department of Energy and the National Science Foundation.
\end{acknowledgments}

\appendix
\section{Discussion of partial waves\label{sec:partialwaves}}
\begin{figure*}[htb]
\begin{center}
\includegraphics[width=0.8\textwidth]{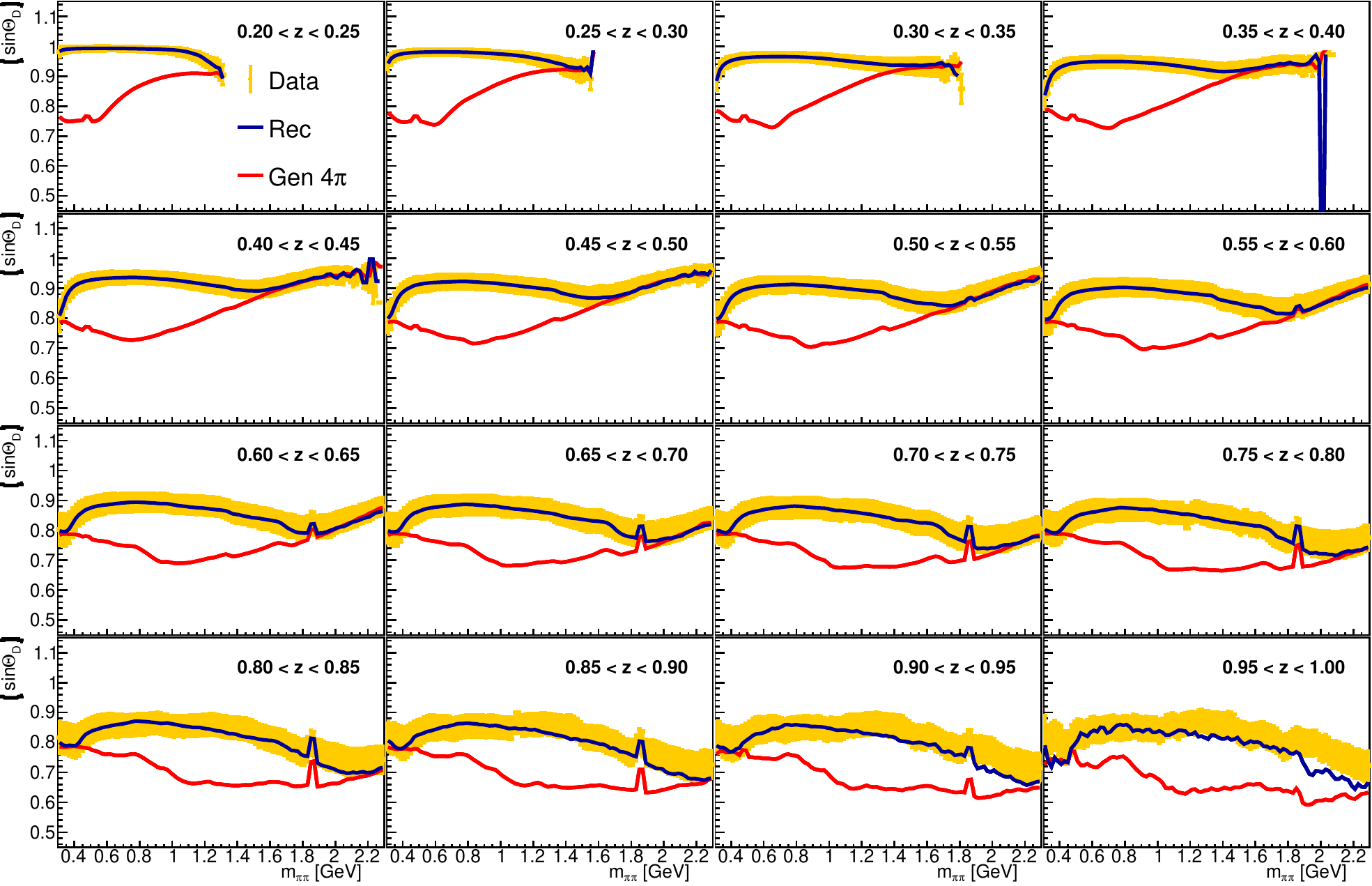}
\caption{\label{fig:sintheta}Average $\sin\theta_D$ decay moments for $\pi^+\pi^-$ pairs as a function of $m_{\pi\pi}$ in bins of $z$. MC on the generator level without minimum momentum requirement (red), after detector simulation and reconstruction (violet) and PID corrected data (yellow) are shown.}
\end{center}
\end{figure*}

\begin{figure*}[hb]
\begin{center}
\includegraphics[width=0.8\textwidth]{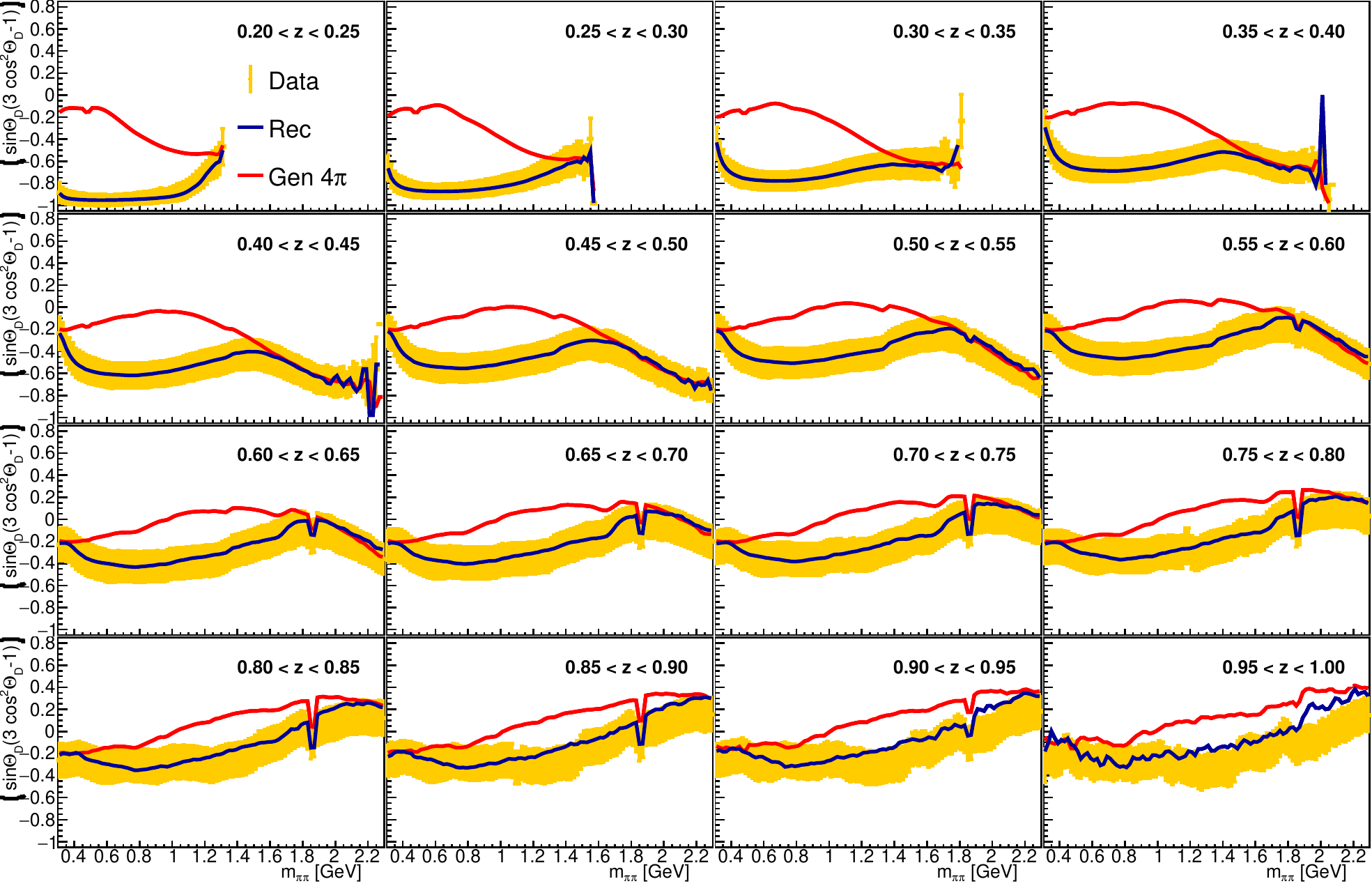}
\caption{\label{fig:sincostheta}Average $\sin\theta_D\left(3\cos^2 \theta_D -1\right)$ decay moments for $\pi^+\pi^-$ pairs as a function of $m_{\pi\pi}$ in bins of $z$. MC on the generator level without minimum momentum requirement (red), after detector simulation and reconstruction (violet) and PID corrected data (yellow) are shown.}
\end{center}
\end{figure*}

As stated in Ref.~\cite{Bacchetta:2002ux}, different partial waves can in principle participate with different strengths. This could be very relevant for the polarized fragmentation function and alter the magnitude of the transversity distributions obtained. For the unpolarized di-hadron cross sections, the corresponding moments of the decay angle, $\theta_D$, relative to the two-hadron direction in its center-of-mass system are studied. Here, the minimal momentum selection removes phase-space for very forward and backward decay angles and prefers decays perpendicular to the di-hadron momentum direction. Comparing the PID corrected (but not unfolded) sine moments of the data to MC at the generator level without a minimum momentum selection, one does see slight increases in the moments. These moments, however, are consistent with a fully tracked and reconstructed MC where such a momentum selection is in place. Again, the $\sin\theta_D\cos\theta_D$ moments of either set of MC and data are consistent with zero while the leading s- and p-wave sine modulations are close to unity. An example of the sine moments for unlike-sign pion pairs is shown in Fig.~\ref{fig:sintheta}. However, in contrast to the asymmetry analysis, the next term in the partial-wave expansion, corresponding to a $\sin\theta_D\left(3\cos^2 \theta_D -1\right)$ moment, does appear to be substantial for pion pairs and even becomes nearly maximal for small fractional energies. An example can be seen in Fig.~\ref{fig:sincostheta}, where it also becomes clear that, at least at lower invariant masses, the magnitude of this moment is dominated by the hadron minimum momentum requirements.
For the polarized di-hadron analysis in Ref.~\cite{belleiff}, one can thus expect a negligible contribution from the p-p interference term of the spin-dependent DiFFs, i.e., to the numerator of Eq.(2) in Ref.~\cite{belleiff}, due to its $\sin\theta_D\cos\theta_D$ modulation \cite{Bacchetta:2002ux}, attributing the nonvanishishing values of the asymmetries to the s-p interference.

\section{Weak decay removed cross sections\label{sec:weakdecays}}
\begin{figure*}[htb]
\begin{center}
\includegraphics[width=0.8\textwidth]{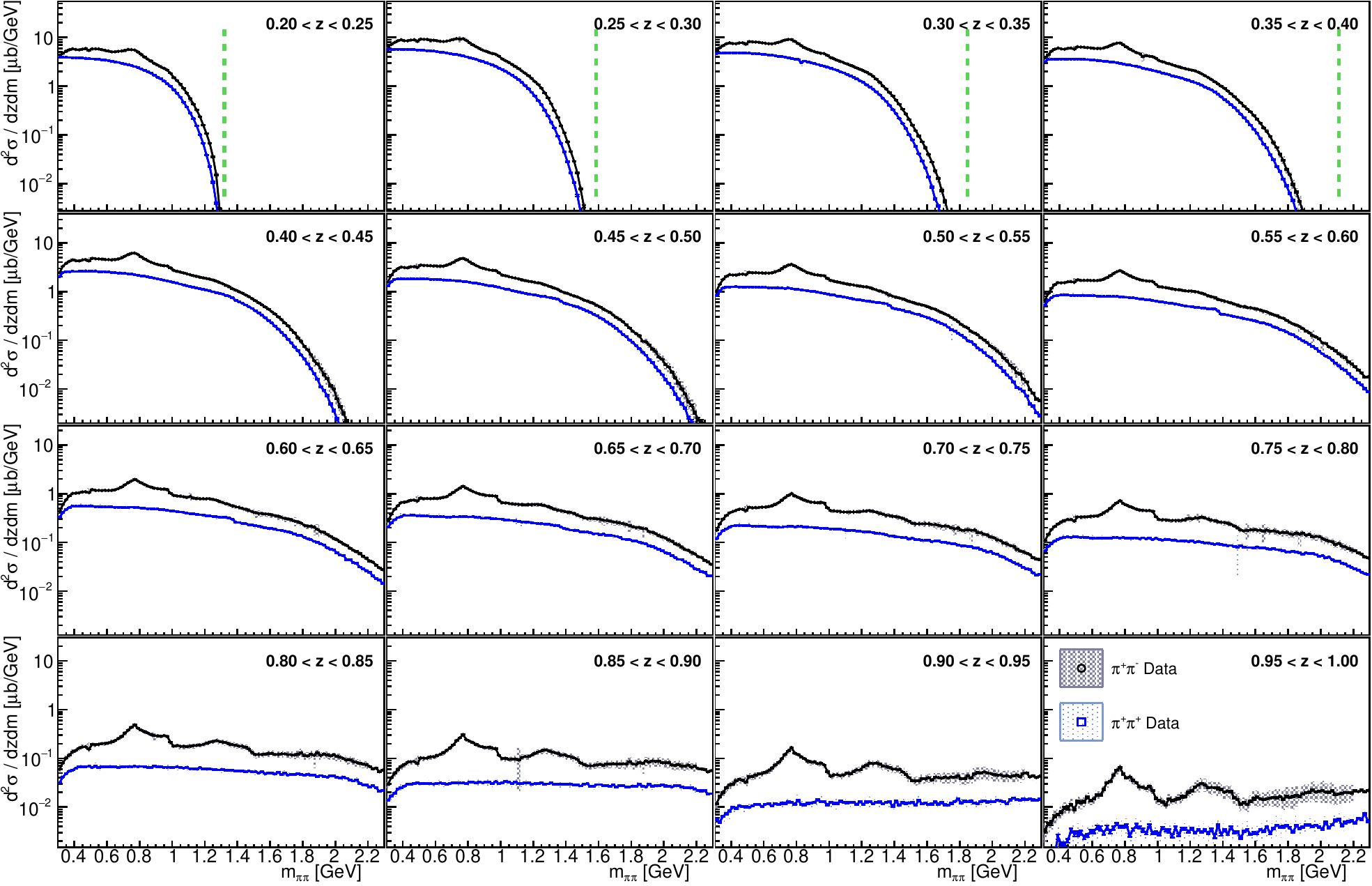}

\caption{\label{fig:allxsec_mix1_pid3strong}Differential cross sections for $\pi^+\pi^-$ (black circles) and $\pi^+\pi^++\ c.c.$ (blue squares) after weak decay removal as a function of $m_{\pi\pi}$ for the indicated $z$ bins. The error boxes represent the systematic uncertainties. The vertical green dashed line corresponds to the kinematic limit. An overall 1.6\% scale uncertainty is not shown.}
\end{center}
\end{figure*}

\begin{figure*}[htb]
\begin{center}
\includegraphics[width=0.8\textwidth]{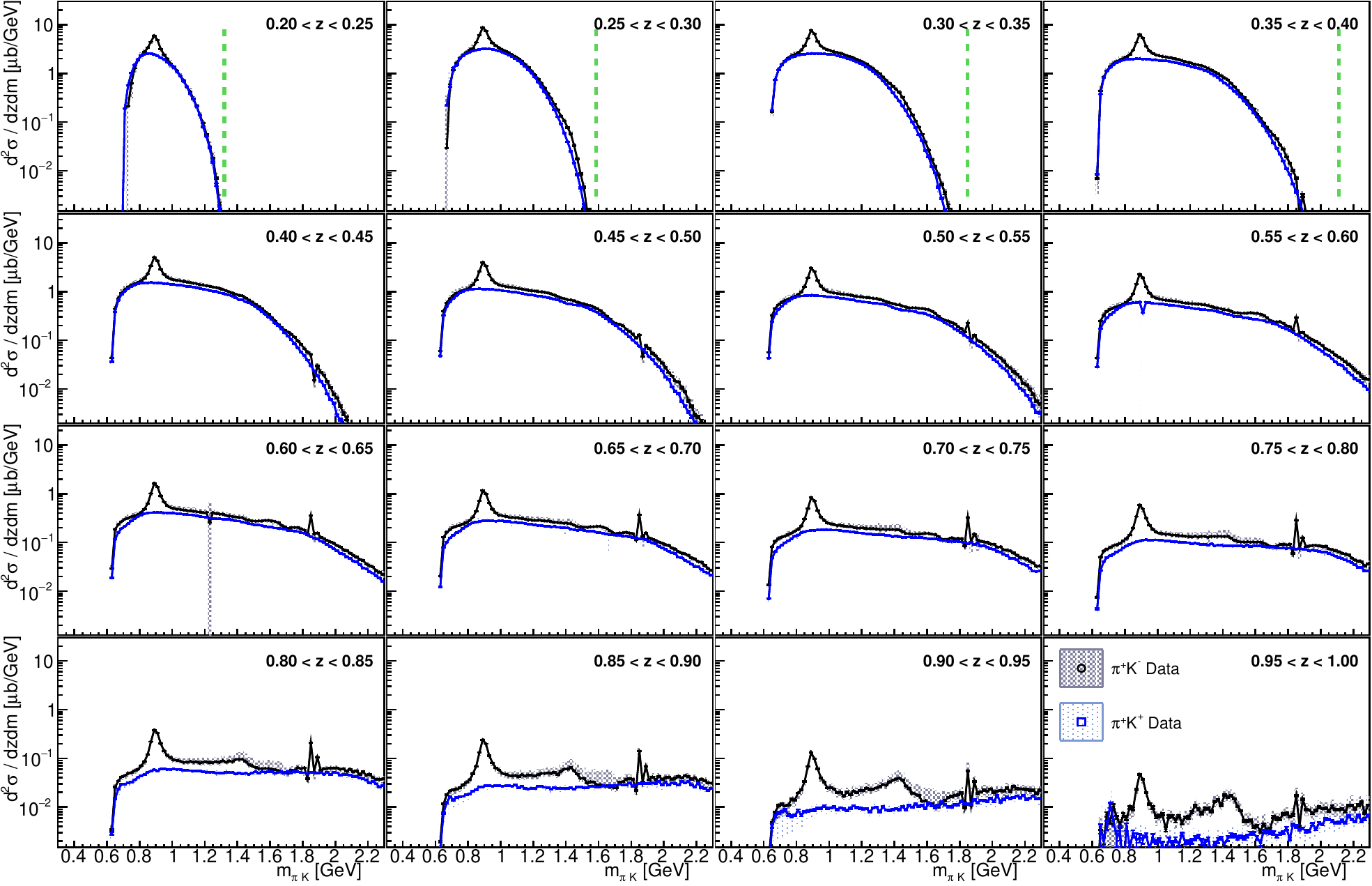}

\caption{\label{fig:allxsec_mix1_pid48strong}Differential cross sections for $\pi^+K^-+\ c.c.$ (black circles) and $\pi^+K^++\ c.c.$ (blue squares) after weak decay removal as a function of $m_{\pi K}$ for the indicated $z$ bins. The error boxes represent the systematic uncertainties. The vertical green dashed line corresponds to the kinematic limit. An overall 1.6\% scale uncertainty is not shown.}
\end{center}
\end{figure*}

\begin{figure*}[hbt]
\begin{center}
\includegraphics[width=0.8\textwidth]{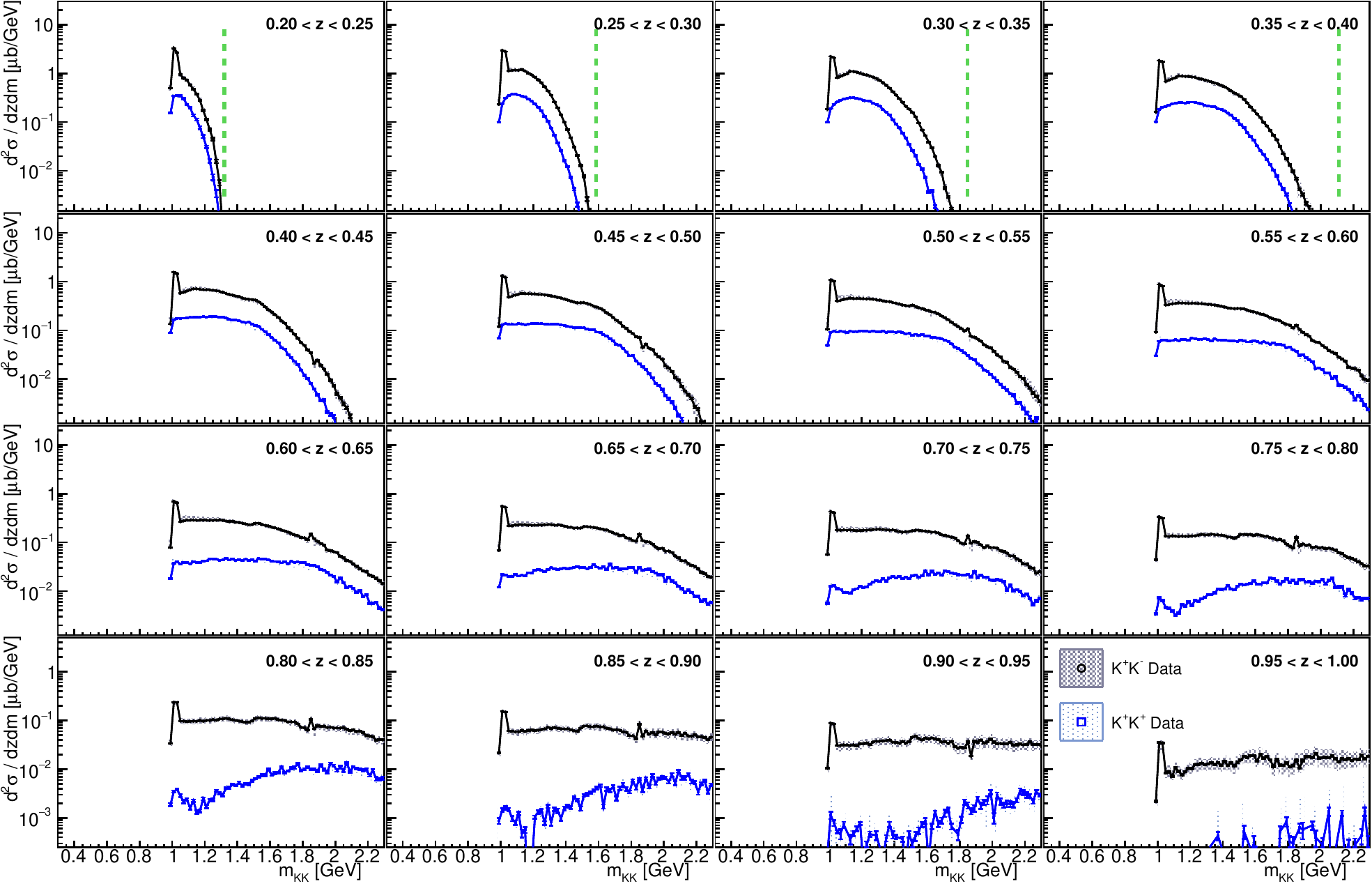}

\caption{\label{fig:allxsec_mix1_pid12288strong}Differential cross sections for $K^+K^-$ (black circles) and $K^+K^++\ c.c.$ (blue squares) after weak decay removal as a function of $m_{KK}$ for the indicated $z$ bins. The error boxes represent the systematic uncertainties. The vertical green dashed line corresponds to the kinematic limit. An overall 1.6\% scale uncertainty is not shown.}
\end{center}
\end{figure*}

In Figs.~\ref{fig:allxsec_mix1_pid3strong}-\ref{fig:allxsec_mix1_pid12288strong}, the di-hadron cross sections after subtraction of hadron pairs from weak decays are displayed. As noted previously, this is based on MC information and can only be as good as the MC description of the overall cross sections. The tune used in this estimation is the Pythia default tune, with the variation due to the tune added as part of the systematic uncertainties in addition to the uncertainties previously discussed. The di-pion cross sections are displayed in Fig.~\ref{fig:allxsec_mix1_pid3strong}. Most notably, the $K^0$ and $D^0$ resonances are not visible anymore while the strongly decaying $\rho^0$ resonance is still prominent. Similarly, in Fig.~\ref{fig:allxsec_mix1_pid48strong} for the pion-kaon pairs, the $K^*$ is visible while the $D^0$ is now missing except for some residual fluctuations around its mass. Apart from the strongly decaying resonances, the same- and opposite-sign pion-kaon pair cross sections are nearly of the same magnitude. 
Lastly, for kaon pairs, only the $\phi$ resonance is visible as can be seen in Fig.~\ref{fig:allxsec_mix1_pid12288strong}.

\end{document}